%% file: perform.tex
\documentclass[namedreferences]{solarphysics}
\usepackage[hyperref,optionalrh,solaromanenum]{spr-sola-addons} % For Solar Physics 
\usepackage{graphicx}                    % For eps figures, newer & more powerfull
\usepackage{color}                       % For color text: \color command
\usepackage{breakurl}                         % For breaking URLs easily trough lines
\usepackage[position=top]{subfig}
\usepackage{booktabs}
\usepackage{siunitx}

\begin{document}

\begin{article}

\begin{opening}

\title{Performance of the Birmingham Solar-Oscillations Network
  (BiSON)}

%%%%%%%%%%%%%%%%%%%%%%%%%%%%%%%%%%%%%%%%%%%%%%%%%%%
%% Authors Names
%
% \author[addressref={},corref,email={}]{\inits{}\fnm{}\lnm{}}

\author[addressref={1},corref,email={s.j.hale@bham.ac.uk}]{\inits{S.J.}\fnm{S.J.}~\lnm{Hale}}
\author[addressref={1}]{\inits{R.}\fnm{R.}~\lnm{Howe}}
\author[addressref={1}]{\inits{W.J.}\fnm{W.J.}~\lnm{Chaplin}}
\author[addressref={1}]{\inits{G.R.}\fnm{G.R.}~\lnm{Davies}}
\author[addressref={1}]{\inits{Y.P.}\fnm{Y.P.}~\lnm{Elsworth}}

%%%%%%%%%%%%%%%%%%%%%%%%%%%%%%%%%%%%%%%%%%%%%%%%%%%
%% Runningheads
%
%\runningauthor{}
%\runningtitle{}

\runningauthor{S.J. Hale {\it et al.}}
\runningtitle{Performance of the Birmingham Solar-Oscillations Network (BiSON)}

%%%%%%%%%%%%%%%%%%%%%%%%%%%%%%%%%%%%%%%%%%%%%%%%%%%
%% Affilations 
%% id shold be the same with \author addressref value.
%\address[id={}]{}

\address[id={1}]{School of Physics and Astronomy, University of Birmingham,
Edgbaston, Birmingham B15 2TT, United Kingdom}

%%%%%%%%%%%%%%%%%%%%%%%%%%%%%%%%%%%%%%%%%%%%%%%%%%%
%%% Abstract 
%\begin{abstract}
%\end{abstract}

\begin{abstract}
The \emph{Birmingham Solar-Oscillations Network} (BiSON) has been
operating with a full complement of six stations since 1992.  Over 20
years later, we look back on the network history. The meta-data from
the sites have been analysed to assess performance in terms of site
insolation, with a brief look at the challenges that have been
encountered over the years.  We explain how the international
community can gain easy access to the ever-growing dataset produced by
the network, and finally look to the future of the network and the
potential impact of nearly 25 years of technology miniaturisation.
\end{abstract}

%%%%%%%%%%%%%%%%%%%%%%%%%%%%%%%%%%%%%%%%%%%%%%%%%%%
%% Keywords
%
%\keywords{}


\end{opening}
%-------------------------------------------------

%%%%%%%%%%%%%%%%%%%%%%%%%%%%%%%%%%%%%%%%%%%%%%%%%%%
%% Sections
%
% \section{}%\label{s:?} 

\input{introduction.tex}
\input{history.tex}
\input{systems.tex}
\input{metrics.tex}
\input{sites.tex}
\input{network.tex}
\input{opendata.tex}
\input{future.tex}

%% Figure 
%
% \begin{figure} 
% \centerline{\includegraphics[width=0.5\textwidth,clip=]{<fig.eps>}}
% \caption{}%\label{fig:?}
% \end{figure}

%% Table
%
% \begin{table}
% \caption{}%\label{tbl:?}
% \begin{tabular}{}     
% \hline
% \multicolumn{2}{c}{<>}
% <data>
% \hline
% \end{tabular}
% \end{table}

%%%%%%%%%%%%%%%%%%%%%%%%%%%%%%%%%%%%%%%%%%%%%%%%%%%%%%%%%%%%%%%%%%%%%%%%%%%
%% Appendix
%
% \appendix   

%%%%%%%%%%%%%%%%%%%%%%%%%%%%%%%%%%%%%%%%%%%%%%%%%%%%%%%%%%%%%%%%%%%%%%%%%%%
%% Acknowledgements
%
% \begin{acks}
%
% \end{acks}

\begin{acks}
We would like to thank all those who are, or have been, associated
with {BiSON}.
In Birmingham: George Isaak, Bill Brookes, Bob van der Raay, Clive
{McLeod}, Roger New, Sarah Wheeler, Clive Speake, Brek Miller, Richard
Lines, Phil Pavelin, Barry Jackson, Hugh Williams, Joe Litherland, Ian
Barnes, Richard Bryan, and John Allison.
In Mount Wilson: Ed Rhodes, Stephen Pinkerton, the team of {USC}
undergraduate observing assistants, former {USC} staff members Maynard
Clark, Perry Rose, Natasha Johnson, Steve Padilla, and Shawn Irish,
and former {UCLA} staff members Larry Webster and John Boyden.
In Las Campanas: Patricio Pinto, Andres Fuentevilla, Emilio Cerda,
Frank Perez, Marc Hellebaut, Patricio Jones, Gast{\'o}n Gutierrez,
Juan Navarro, Francesco Di Mille, Roberto Bermudez, and the staff of
{LCO}.
In Iza\~na: We would like to give particular thanks to Pere
{Pall{\'e}} and Teo {Roca Cort{\'e}s}, and all staff at the {IAC} who
have contributed to running the {Mark~I} instrument over many years
(see also the acknowledgements in~\opencite{2014MNRAS.443.1837R}).
In Sutherland: Pieter Fourie, Willie Koorts, Jaci Cloete, Reginald
Klein, John Stoffels, and the staff of {SAAO}.
In Carnarvon: Les Bateman, Les Schultz, Sabrina Dowling-Giudici, Inge
Lauw of Williams and Hughes Lawyers, and {NBN} Co.\ Ltd.
In Narrabri: Mike Hill and the staff of {CSIRO}.
{BiSON} is funded by the Science and Technology Facilities Council
({STFC}).
\end{acks}

%%%%%%%%%%%%%%%%%%%%%%%%%%%%%%%%%%%%%%%%%%%%%%%%%%%%%%%%%%%%%%%%%%%%%%%%%%%
%% CoI
%

\section*{Disclosure of Potential Conflicts of Interest}
The authors declare that they have no conflicts of interest.

%%% %%%%%%%%%%%%%%%%%%%%%%%%%%%%%%%%%%%%%%%%%%%%%%%%%%%%%%%%%%%
%% Bibliography
%
% Using BibTeX
%
% \bibliographystyle{spr-mp-sola}
% \bibliography{<bib file>}  
%
% Without BibTeX 
% \begin{thebibliography}{}
% \bibitem[\protect\citeauthoryear{Author}{Year}]{key}
%   <bibliographical entry>
%
% \bibitem[\protect\citeauthoryear{}{}]{}
%   
%  
% \end{thebibliography}

\bibliographystyle{spr-mp-sola}
\bibliography{perform}

\end{article} 
\end{document}

%% file: introduction.tex
% INTRODUCTION.TEX
%
%   Steven Hale
%   2015 February 2
%   Birmingham, UK
%
% BiSON Performance Paper
%
% $Id: introduction.tex 59 2015-10-17 06:44:36Z hale $
%

\section{Introduction}
     \label{S-Introduction}

The \emph{Birmingham Solar Oscillations Network} (BiSON) has now been
operating continuously as a six-station network for well over twenty
years, recording high-quality spatially unresolved, or
``Sun-as-a-star'' helioseismic data.  It therefore now seems timely to
update our previous article on BiSON performance
\citep{1996SoPh..168....1C}. We will present updated results on the
temporal coverage and noise performance of the individual sites and
the network as a whole and reflect on what we have learned from more
than two decades of experience in operating a semi-automated
ground-based observing network.  These data are available to the wider
community through the BiSON Open Data Portal.

The early history of the network has been outlined by
\cite{1996SoPh..168....1C}. In brief, early campaign-style
observations from two sites (Haleakala and Iza\~na) were followed by
the addition of an automated observing station in Carnarvon, Western
Australia and later by the deployment of more standardised stations in
Sutherland, South Africa, Las Campanas, Chile, and Narrabri, Australia
over the period~1990\,--\,1992. The instrument from Haleakala was
moved to California and installed in the 60~foot tower at the Mount
Wilson Hale Observatory in~1992. Since then, with occasional
instrument upgrades, the network has been operating continuously,
providing unresolved-Sun helioseismic observations with an average
annual duty cycle of about 82\,\%.

%% Figure 
%
% \begin{figure} 
% \centerline{\includegraphics[width=0.5\textwidth,clip=]{<fig.eps>}}
% \caption{}%\label{fig:?}
% \end{figure}

%% Table
%
% \begin{table}
% \caption{}%\label{tbl:?}
% \begin{tabular}{}     
% \hline
% \multicolumn{2}{c}{<>}
% <data>
% \hline
% \end{tabular}
% \end{table}

%% file: history.tex
% HISTORY.TEX
%
%   Steven Hale
%   2015 July 21
%   Birmingham, UK
%
% Observational helioseismology: a brief history
%
% $Id: history.tex 61 2015-10-22 05:07:45Z hale $
%

%%%%%%%%%%%%%%%%%%%%%%%%%%%%%%%%%%%%%%%%%%%%%%%%%%%%%%%%%%%%%%%%

\section{Observational Helioseismology: A Brief History}
     \label{S-History}

Oscillations in the velocity field across the Sun were first
discovered by Robert Leighton \citep{1962ApJ...135..474L} using the
``spectroheliograph'' developed by George Ellery Hale at the Mount
Wilson Hale Observatory in California.  The now accepted explanation
for these oscillations was developed by Roger Ulrich and John
Leibacher, both independently suggesting that sound waves could be
generated in the convection zone of the solar
interior~\citep{1970ApJ...162..993U,1971ApL.....7..191L}.  Interest in
the new field of helioseismology grew quickly, culminating in a
dataset six days in length collected from the South
Pole~\citep{1980Natur.288..541G}, which at the time was the longest
and most detailed set of continuous observations available.

It soon became clear that long-term continuous observations were
required.  The Birmingham group was the first to begin construction of
a network of ground-based observatories, beginning with {Iza\~na} in
Tenerife in~1975 and culminating in six operational sites in~1992.

Other groups have also had success with ground based networks.  Fossat
and colleagues went on to deploy the \emph{International Research of
  Interior of the Sun} (IRIS;~\opencite{1991SoPh..133....1F}) network.
The operational strategy of {IRIS} was different from {BiSON},
requiring full participation of local scientists responsible for each
instrument as opposed to automation.  The instrumentation operated in
a similar manner to {BiSON}, observing spatially unresolved global low
angular-degree oscillations using an absorption line of sodium, rather
than the potassium line used by {BiSON}.  The {IRIS} network spanned
six sites in total (up to nine sites when considering additional
collaboration~\citep{2002A&A...390..717S}), and was operational
until~2000~\citep{2002ESASP.508...95S,2002sf2a.conf..521F}.

Leibacher and colleagues developed the \emph{Global Oscillation
  Network Group} (GONG;~\opencite{1996Sci...272.1284H}), a six-site
automated network using resolved imaging to observe modes of
oscillation at medium degree (up to $l\approx 150$), complementary to
the existing low-degree networks.  The sites were chosen in
early~1991, and deployment began in~1994 with instruments coming
online throughout~1995.  The GONG network was upgraded in
2001\,--\,2002 to observe up to around $l=1000$, and is still in
operation today.

For completeness, we should also note the LOWL project
\citep{1995ApJ...448L..57T}, which observed at medium degree from one
or two sites between 1994 and 2004, and the \emph{Taiwanese
  Oscillations Network} \citep{1995SoPh..160..237C} for high-degree
observations, which was deployed between 1993 and 1996 but operated
for only a few years.

Several space-based missions have also been successful.  In the early
1980s the \emph{Active Cavity Radiometer Irradiance Monitor}
({ACRIM};~\opencite{1979ApOpt..18..179W}) onboard the NASA \emph{Solar
  Maximum Mission} spacecraft was sufficiently precise to detect the
small changes in intensity caused by the oscillations.  Later, the
\emph{Solar and Heliospheric Observatory} ({SOHO}), a joint project
between {ESA} and {NASA} was launched in December 1995 and began
normal operations in May~1996~\citep{1995SoPh..162....1D}.  Among the
suite of instruments onboard the spacecraft are three helioseismic
instruments: \emph{Global Oscillations at Low Frequencies}
(GOLF;~\opencite{1995SoPh..162...61G}) and \emph{Variability of solar
  Irradiance and Gravity Oscillations}
(VIRGO;~\opencite{1995SoPh..162..101F}) for low-degree oscillations
and \emph{Michelson Doppler Imager}
(MDI;~\opencite{1995SoPh..162..129S}) for observations at medium and
high degree.  The mission was originally planned for just two years,
but it is still in operation and currently has a mission extension
lasting until December~2016, at which point it will have been in
service for over twenty years.  MDI ceased observations in 2011 when
it was superceded by the \emph{Helioseismic and Magnetic Imager}
(HMI;~\opencite{2012SoPh..275..229S}) onboard the \emph{Solar Dynamics
  Observatory}~(SDO), but GOLF and VIRGO are still in use.

%%%%%%%%%%%%%%%%%%%%%%%%%%%%%%%%%%%%%%%%%%%%%%%%%%%%%%%%%%%%%%%%

%% file: systems.tex
% SYSTEMS.TEX
%
%   Steven Hale
%   2015 June 4
%   Birmingham, UK
%
% BiSON Control Systems
%
% $Id: systems.tex 61 2015-10-22 05:07:45Z hale $
%

%%%%%%%%%%%%%%%%%%%%%%%%%%%%%%%%%%%%%%%%%%%%%%%%%%%%%%%%%%%%%%%%

\section{Designing an Automated Robotic Network}
     \label{S-Designing an Automated Robotic Network}

%%%%%%%%%%%%%%%%%%%%%%%%%%%%%%%%%%%%%%%%%%%%%%%%%%%%%%%%%%%%%%%%

The principle of using resonant scattering spectroscopy to achieve
stable and precise measurements of the line-of-sight velocity of the
solar atmosphere was first proposed by~\cite{1961Natur.189..373I}.
The optical design of the instrument first used at Pic-du-Midi in~1974
to detect long-period solar oscillations is described
by~\cite{1976Natur.259...92B}.  The basic observational parameter is
the Doppler shift of the solar potassium Fraunhofer line
at~\SI{770}{\nano\meter}.  This is achieved through measurement of
intensity over a very narrow range in wavelength that sits in the
wings of the potassium line so that intensity changes with Doppler
shift, and through comparison with the same transition in a potassium
vapour in the laboratory the change in intensity is calibrated to
become a measure of velocity.  Following initial tests at Pic-du-Midi,
the spectrometer was then relocated to the Observatorio del Teide,
Tenerife, during 1975.  The updated apparatus is described
by~\cite{1978MNRAS.185....1B}.

The six-station network of today was completed in~1992.  There are two
stations in each~\SI{120}{\degree} longitude band, and all of the
sites lie at moderate latitudes, around~\SI{\pm30}{\degree}.  The
oldest site in the Birmingham Network, Iza\~na, has now been
collecting data for nearly forty years.

The original control system was based around a {40-channel} scaler
module used for counting pulses from a photomultiplier
tube~\citep{bison184}.  Timing was controlled by a quartz clock,
related to {GMT} at least once per day, and producing pulses at
exactly~\SI{1}{\second} intervals.  At the end of each interval two
relays would change state and reverse the voltage across an
electro-optic modulator, and the data gate would be incremented to the
next channel of the scaler.  Once all 40 channels of the scaler had
been filled the contents would be destructively written out to
magnetic tape, a process which required a further~\SI{2}{\second}.
The process would then repeat, with each block of data separated
by~\SI{42}{\second}.  The cadence was changed to~\SI{40}{\second} at
the beginning of~1990.  This allows for simpler concatenation of data
from different sites since there are an integer multiple
of~\SI{40}{\second} in a day, and so it provides a network
time-standard.

The system was computerised in 1984 using the {BBC Microcomputer}.
The {BBC Micro} was originally commissioned on behalf of the {British
Broadcasting Corporation} as part of their {Computer Literacy
Project}, and it was designed and built by the {Acorn Computer}
company.  Despite the {BBC Micro} being discontinued in~1994, the
spectrometer continued operating in this configuration until~2003 when
the computer finally failed and was replaced by a modern PC.  A
dedicated {PIC}-based interface was designed to enable the PC to
communicate with the original scaler system~\citep{bison218,bison243},
and this remains the operating configuration today.

For the subsequent fully automated solar observatories, the
data-acquisition and control system was initially based around a
{Hewlett-Packard 3421A Data Acquisition Unit}, known colloquially as a
``data logger''.  By the early 1990s these were retired from service
and replaced with a standard desktop personal computer and a Keithley
System~570 digital input/output interface.  The computer ran Microsoft
DOS.  The limitations of the operating system meant that data could
only be retrieved by Birmingham during a scheduled ``window'' when the
computer would switch from running the data-acquisition program to
running a data-transfer program.  Over the years a variety of means
have been used to return the data from the sites to Birmingham,
ranging from tape cartridges or floppy disks sent by post, through
direct dial-up modem connections over international phone lines, to
the modern internet.

Initially, the instrument at the first automated dome in Carnarvon
observed through a glass window.  Although the concern regarding site
security was low, the window was considered necessary for safe
operation.  Despite being cleaned regularly by the on-site support,
marks on the glass were detrimental to the data and the decision was
eventually made to remove the window and operate in the conventional
style of a simple shutter opening.  This meant that since the whole
system is designed to run completely unattended, careful weather
monitoring was required.  Both rain and wind sensors are used to close
the dome in the event of precipitation or excessive wind.  All of the
subsequent observatories operated without window glass.

Naturally over the years repairs have been made and upgrades
installed.  Probably the most significant is the upgrade to the
control software.  The original Keithley System~570 data-acquisition
system in Carnarvon ran for more than a decade of continuous use, but
the units eventually became obsolete as PCs failed and the
replacements lacked the required {ISA} interface. At the same time,
the {MS-DOS} operating system on which the original dome control
software relied itself became obsolete, and the Windows software that
superseded it for home and office uses was unsuitable for system
automation. A new dome control system known as the ``Zoo'' was
developed in the late 1990s~\citep{bison187,bison193} to run under
the recently-released {GNU/Linux} operating system~\citep{gnu,linux},
and this continues in use to the present day.

Much of the old analogue electronics have been gradually replaced with
new digital variants.  Rather than the whole system consisting of
modules in a central rack and communicating through a single
interface, the new designs are independent with embedded
micro-controllers and communicate with the PC through dedicated
{RS-232} serial ports.  This makes subsequent repairs and upgrades
considerably easier since units can be inspected in isolation.

In the following sections we look back over the performance of the
network, including temporal coverage and noise levels, and also
discuss some of the significant events during the life of the
stations.

%%%%%%%%%%%%%%%%%%%%%%%%%%%%%%%%%%%%%%%%%%%%%%%%%%%%%%%%%%%%%%%%

%% file: metrics.tex
% METRICS.TEX
%
%   Steven Hale
%   2015 April 17
%   Birmingham, UK
%
% BiSON Data Quality Metrics
%
% $Id: metrics.tex 59 2015-10-17 06:44:36Z hale $
%

%%%%%%%%%%%%%%%%%%%%%%%%%%%%%%%%%%%%%%%%%%%%%%%%%%%%%%%%%%%%%%%%

\section{Data Quality Metrics}
     \label{S-Data Quality Metrics}

Before we can discuss data quality, we must first define some quality
metrics.  There are two standard metrics used by {BiSON}.  These are
the \emph{five-minute figure of merit} known simply as the {FOM}, and
the mean high-frequency noise level.

Data from {BiSON} are collected on a cadence of~\SI{40}{\second},
giving an upper limit in the frequency domain (Nyquist frequency) of
\SI{12.5}{\milli\hertz}.  The {FOM} is a signal-to-noise ratio, and is
defined as the total power in the main ``five-minute'' signal
band~(\SI{2}{\milli\hertz}\,--\,\SI{5}{\milli\hertz}) divided by the
noise~(\SI{5.5}{\milli\hertz}\,--\,\SI{12.5}{\milli\hertz}).  When
considering the mean noise level in isolation, we look at the mean
power in the high-frequency
noise~(\SI{10.0}{\milli\hertz}\,--\,\SI{12.5}{\milli\hertz}).  We will
refer to these definitions of \emph{FOM} and \emph{mean noise}
throughout this article.

\begin{figure}%
    \centering
    \subfloat[]{\includegraphics[width=0.47\textwidth]{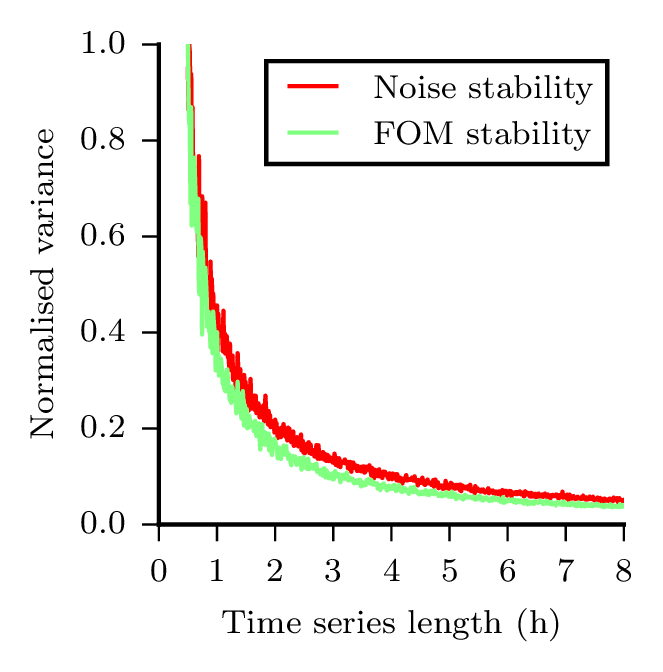}\label{fig:fom}}%
    \qquad
    \subfloat[]{\includegraphics[width=0.47\textwidth]{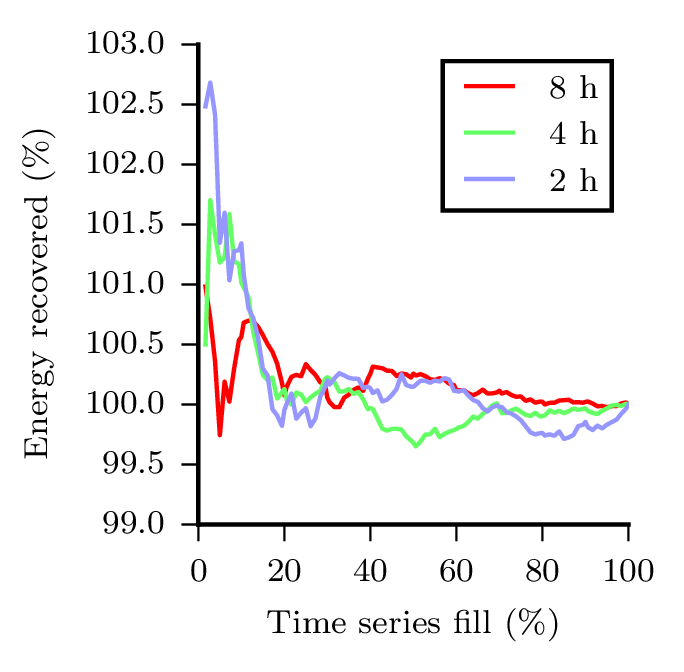}\label{fig:fill}}%
    \caption{Measured FFT stability over varying dataset length and
      fill.  For a valid comparison between datasets from different
      days and different sites, a time series needs to be at least
      ~\SI{3}{\hour} in duration and have a fill of at least 25\,\%.}
    \label{fig:fftstability}%
\end{figure}

Both of these metrics require a certain volume of data in order to
ensure that the quality estimates are reliable, and that meaningful
comparisons can be made between data from different days and different
instruments.  As the length of a dataset decreases, different
realisations of random noise begin to have a larger impact on the
estimate of quality.  Eventually the quality estimate becomes so
variable as to be meaningless.  In order to determine the minimum
dataset length required to make reliable comparisons, a simple
artificial dataset was produced with one thousand realisations of
random noise.  The FOM and mean noise were calculated for each
realisation, and the variance in the FOM and mean noise recorded.
This was done for dataset lengths varying from thirty minutes up to
eight hours.  The normalised results are shown in
Figure~\ref{fig:fftstability}\subref{fig:fom}.  Both the quality
metrics appear to stabilise at dataset lengths of a minimum
of~\SI{3}{\hour}, and so this was selected as the minimum length that
could be reliably used for quality comparison in this article.

An additional problem with data from BiSON is that due to variable
weather conditions it is usually not continuous.  In order to be able
to compare absolute noise level we need to rescale the power spectrum
produced by the FFT to compensate for any missing data.  This is done
by simply dividing by the percentage fill (\textit{i.e.} the
percentage of the total observing time where data are available).  To
determine how low the fill can be whilst still providing a meaningful
estimate of the total energy in an equivalent gap-free dataset, a
similar test was used with simple artificial data and one thousand
realisations of random noise.  A variable size gap was created in
three datasets of two, four, and eight hours in total length.  The
results for percentage of total energy recovered compared with the
original time series are shown in
Figure~\ref{fig:fftstability}\subref{fig:fill}.  Even with very low
fills it is possible to recover an estimate of the original total
energy to within a few percent.  For this article, a fill of at least
25\,\% was selected as being the minimum requirement.

%%%%%%%%%%%%%%%%%%%%%%%%%%%%%%%%%%%%%%%%%%%%%%%%%%%%%%%%%%%%%%%%

%% file: sites.tex
% SITES.TEX
%
%   Steven Hale
%   2015 April 17
%   Birmingham, UK
%
% BiSON Statistics - Sites
%
% $Id: sites.tex 62 2015-10-30 03:46:22Z hale $
%

%%%%%%%%%%%%%%%%%%%%%%%%%%%%%%%%%%%%%%%%%%%%%%%%%%%%%%%%%%%%%%%%

\section{Site Performance}
     \label{S-Site Performance}

%%%%%%%%%%%%%%%%%%%%%%%%%%%%%%%%%%%%%%%%%%%%%%%%%%%%%%%%%%%%%%%%

\subsection{Iza\~na, Tenerife}

As we saw at the beginning of Section~\ref{S-Designing an Automated
  Robotic Network}, the instrument that was to become the first
``node'' of the \emph{Birmingham Solar Oscillations Network}
({BiSON}), was installed at the Observatorio del Teide, Tenerife,
in~1975.  The Birmingham group were the first to establish a global
network of ground-based observatories dedicated to helioseismology.

George Isaak, the then head of the \emph{High-Resolution Optical
  Spectroscopy} ({HiROS}) research group, was already considering the
development of a permanent network to expand beyond Tenerife long
before his seminal article in 1979 on global studies of the
five-minute oscillation~\citep{1979Natur.282..591C}.  Campaign-style
operations continued throughout~1978 and~1979 at Pic-du-Midi in the
French Pyrenees, and a short run at Calar Alto in Spain in~1980.  In
1981 the group secured funding to operate a second site on the island
of Haleakala in Hawaii, at the Mees Observatory.  The instrument at
Haleakala operated in much the same way as Mark~I in Tenerife.  From
the two sites together data were collected for 88~days, producing the
longest time series, and most highly resolved power spectrum, that had
been achieved up to that time.  However, this was still a long way
from year-round complete coverage.

Mark~I is housed in the ``Pyramid Building'' at the Observatorio del
Teide, and run by Pere Pall\'e and his team of
observers~\citep{2014MNRAS.443.1837R}.  Light is collected via two
mirrors, known as a c{\oe}lostat.  The beam is projected through an
open window into the apex of the pyramid, where Mark~I sits on an
optical bench.  At the beginning of each observing session, the
on-site observer needs to uncover and align the mirrors and start the
system.  Operator presence is also required throughout the day to
close the mirrors in the event of bad weather, and at the end of the
observing session.

\begin{figure}
\centering
\includegraphics[width=0.9\textwidth,height=0.9\textheight,keepaspectratio]{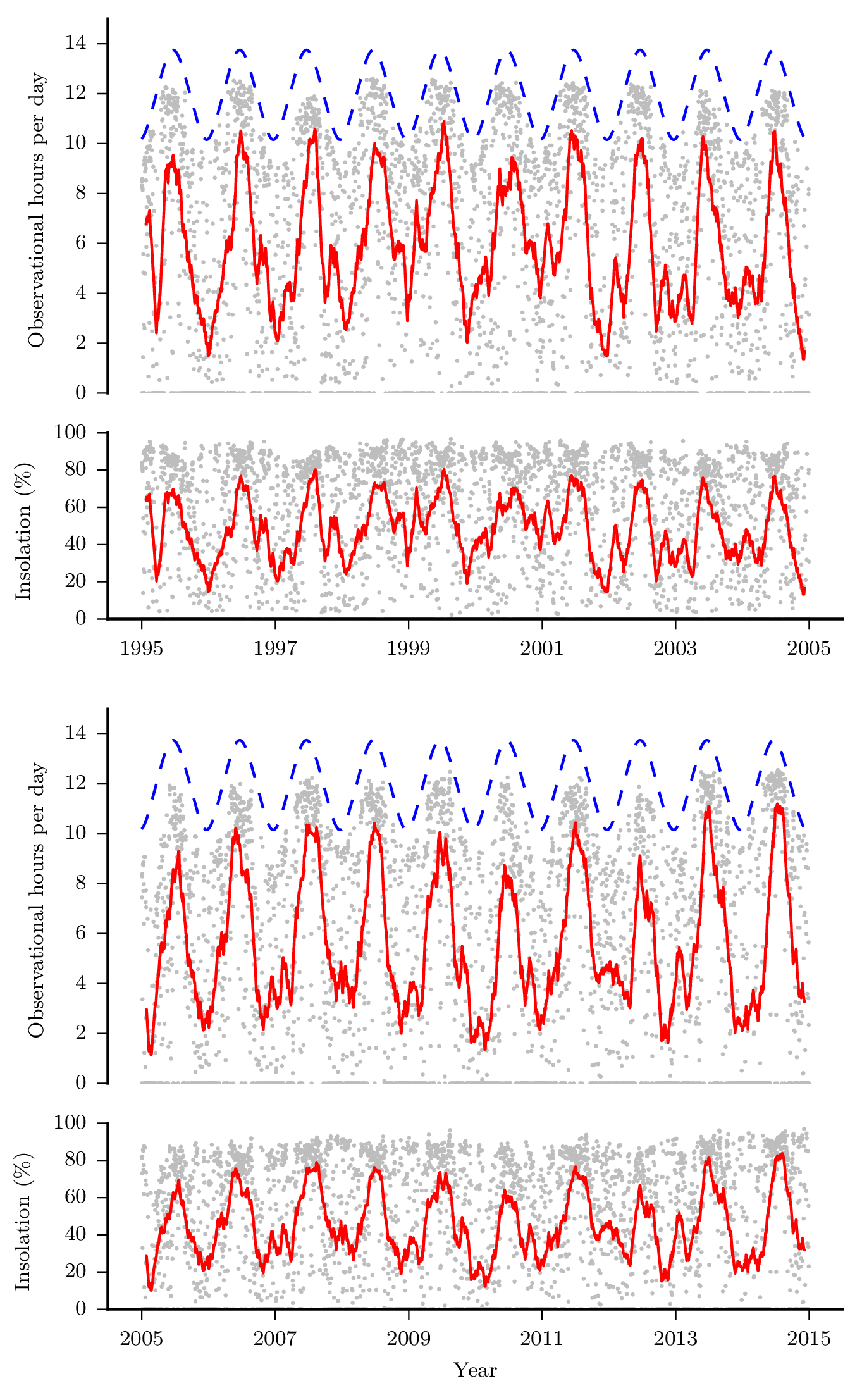}
\caption{Iza\~na duty cycle as a function of date, plotted in hours
  per day, and as a percentage of potential daylight hours. There is
  one grey dot per day, and the solid red curve represents a 50-day
  moving mean.  The dashed blue curve shows potential daylight hours.}
\label{fig:izana_fill}
\end{figure}

Despite the higher photon shot noise level compared to our more modern
sites, Iza\~na has been and still is a work-horse of the {BiSON}
network.  The site duty-cycle is shown in Figure~\ref{fig:izana_fill}.
The duty-cycle is plotted in terms of both number of observational
hours, and also percentage insolation.  The insolation compares only
the potential daylight hours against actual observational hours, and
does not differentiate between poor weather conditions and any periods
of instrumental failure.  Tenerife provides exceptional weather
conditions over the summer months, but is rather poor throughout the
winter where the conditions become much more variable.  There are
regular ``holes'' in the site window function at midday during the
spring and autumn months which are caused by c{\oe}lostat shadowing.
The secondary mirror has two mounting positions, from above or from
below, which allows an unobstructed daily run to be achieved
throughout the winter and summer months.  However, during the
change-over period between the two phases the secondary mirror
unavoidably shadows the primary.

\begin{figure}
\centering \includegraphics[width=0.9\textwidth,height=0.9\textheight,keepaspectratio]{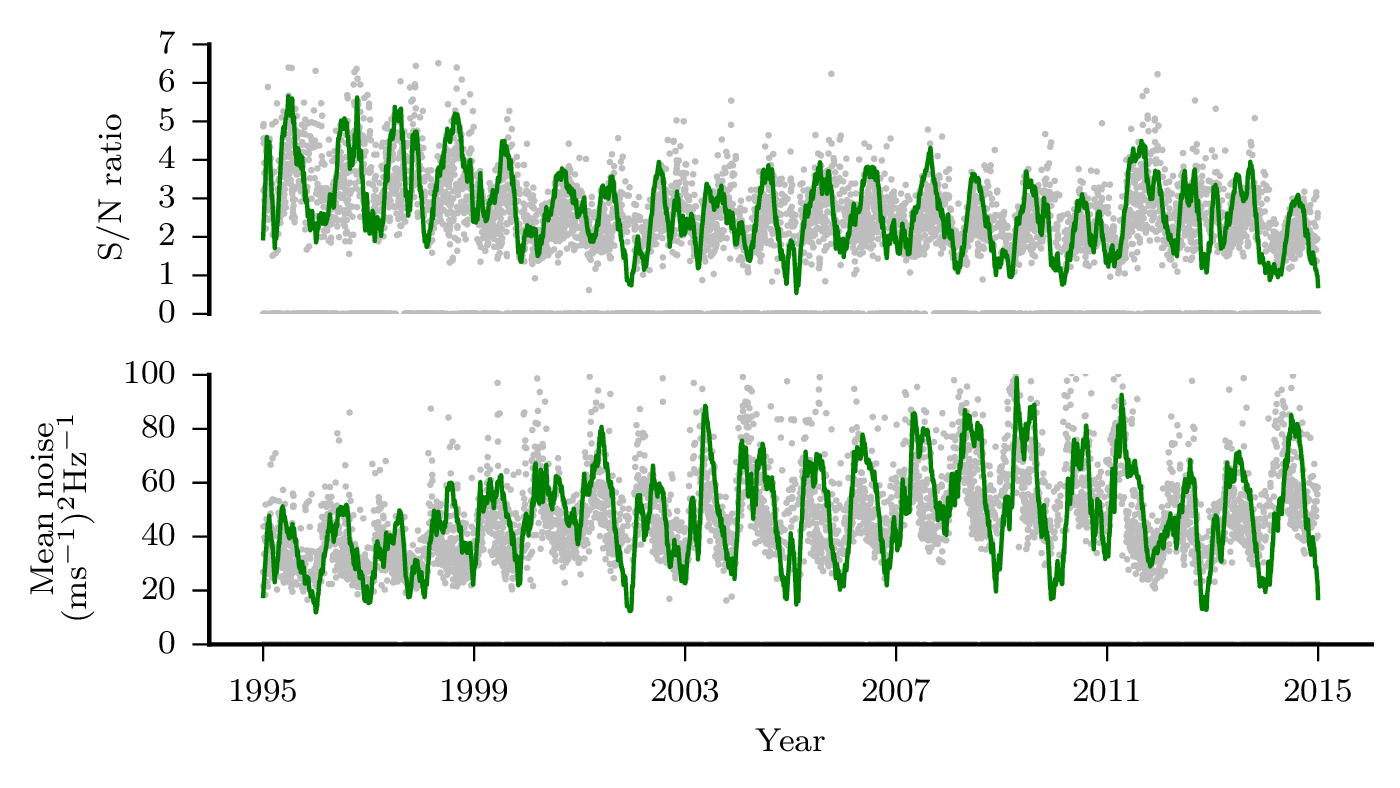}
\caption{Iza\~na data quality as a function of date.  Top: Signal
  to noise ratio, higher is better.  Bottom: Mean noise level, lower
  is better.  There is one grey dot per day, and the solid green curve
  represents a 50-day moving mean.}
\label{fig:izana_quality}
\end{figure}

The figure of merit ({FOM}) and mean noise levels for Iza\~na are
shown in Figure~\ref{fig:izana_quality}.  The seasonal variation in
noise level, and subsequent change in {FOM}, is due to an effect
described by~\cite{2004ESASP.559..360C,2005MNRAS.359..607C}.  The
basic measurement of a {BiSON} {RSS} is that of intensity change due
to the shift of a solar Fraunhofer line.  As the Fraunhofer line
shifts due to the line-of-sight relative motion of the solar surface
with respect to the laboratory, the measured intensity changes.  A
data pipeline calibrates the intensity measurements into velocity,
taking into account the non-linearity of the Fraunhofer line shape.
However, the propagation of noise on the original intensity
measurement through the calibration process is
non-linear~\citep{1989ApJ...345.1088H} due to the varying types of
noise.  The system encounters white noise, for example due to photon
statistics and analogue-to-digital conversion, multiplicative noise
due to gain fluctuations, and additive noise due to offset
fluctuations, in addition to more random effects due to pointing
errors and temperature fluctuations.  As the Fraunhofer line shifts
daily due to Earth's rotation, and seasonally due to the eccentricity
of Earth's orbit, the line gradient at the operating point of the
instrument changes.  This causes the noise level in the derived
quantities to change even if the noise level in the basic intensity
measurement remains constant.  Aside from the seasonal variation the
noise levels from Iza\~na show remarkable stability and offer an
unprecedented temporal baseline of almost 40~years.

%%%%%%%%%%%%%%%%%%%%%%%%%%%%%%%%%%%%%%%%%%%%%%%%%%%%%%%%%%%%%%%%

\subsection{Carnarvon, Western Australia}

When considering expanding to a global network the group realised that
it would not be practical to operate such a network manually.
Requiring an observer to be present on site, all day every day for
365~days per year, would be very expensive and potentially unreliable.
As the number of network nodes increases, the number of observers
required to operate them all year round would become untenable.  The
key to a reliable network would be automation.  For the early 1980s,
this was an ambitious endeavour.  If successful, it would be one of
the first automated astronomical telescopes anywhere in the world, in
any field.

From~1981 the group worked on this innovative new design, and by the
summer of~1983 a prototype was ready to test.  The new instrument
would point directly at the Sun, and move on an equatorial mount like
a classical astronomical telescope.  Moving a mount under computer
control would be significantly easier than aligning and pointing
mirrors.  Testing in Haleakala proved that the system worked, and the
group began looking for an installation site.  Western Australia was
selected as a good longitude to complement coverage from Pic-du-Midi
and Haleakala.  Four sites were investigated: Woomera Rocket Testing
Range, Learmonth, Exmouth, and Carnarvon.  Isaak decided that Woomera
was too dusty, but the other three towns were all suitable.  The group
settled on Carnarvon, some~\SI{900}{\kilo\metre} north of Perth.  The
group arrived in Carnarvon in~1984.  Similar testing was carried out
to that which had been successful in Haleakala.  By the end of the
year it was clear that full automation was practical, and the decision
was made to push ahead.

Initially the data collection systems operated on a~\SI{42}{\second}
cadence like the existing systems in Iza\~na and Haleakala.  Carnarvon
was migrated to the newer~\SI{40}{\second} cadence in April~1992.
Data obtained previous to this date are interpolated onto the newer
standard cadence.  Over the months following installation, the system
worked extremely well.  It produced high-quality data and proved to be
reliable.  Most glitches were due to the Australian wildlife --
rodents eating through cables, or cockatoos making nests in the upper
parts of the dome.  The Sun itself caused some problems, with certain
types of connector and cable insulation quickly decaying under
constant exposure to solar~UV.  A simple and cheap solution was found
for this, which was to wrap exposed cables and components in aluminium
cooking foil.  The goal of demonstrating that an automated system
could work, and work well, had been successfully achieved.  Funding to
roll out further stations was quickly secured.

\begin{figure}
\centering
\includegraphics[width=0.9\textwidth,height=0.9\textheight,keepaspectratio]{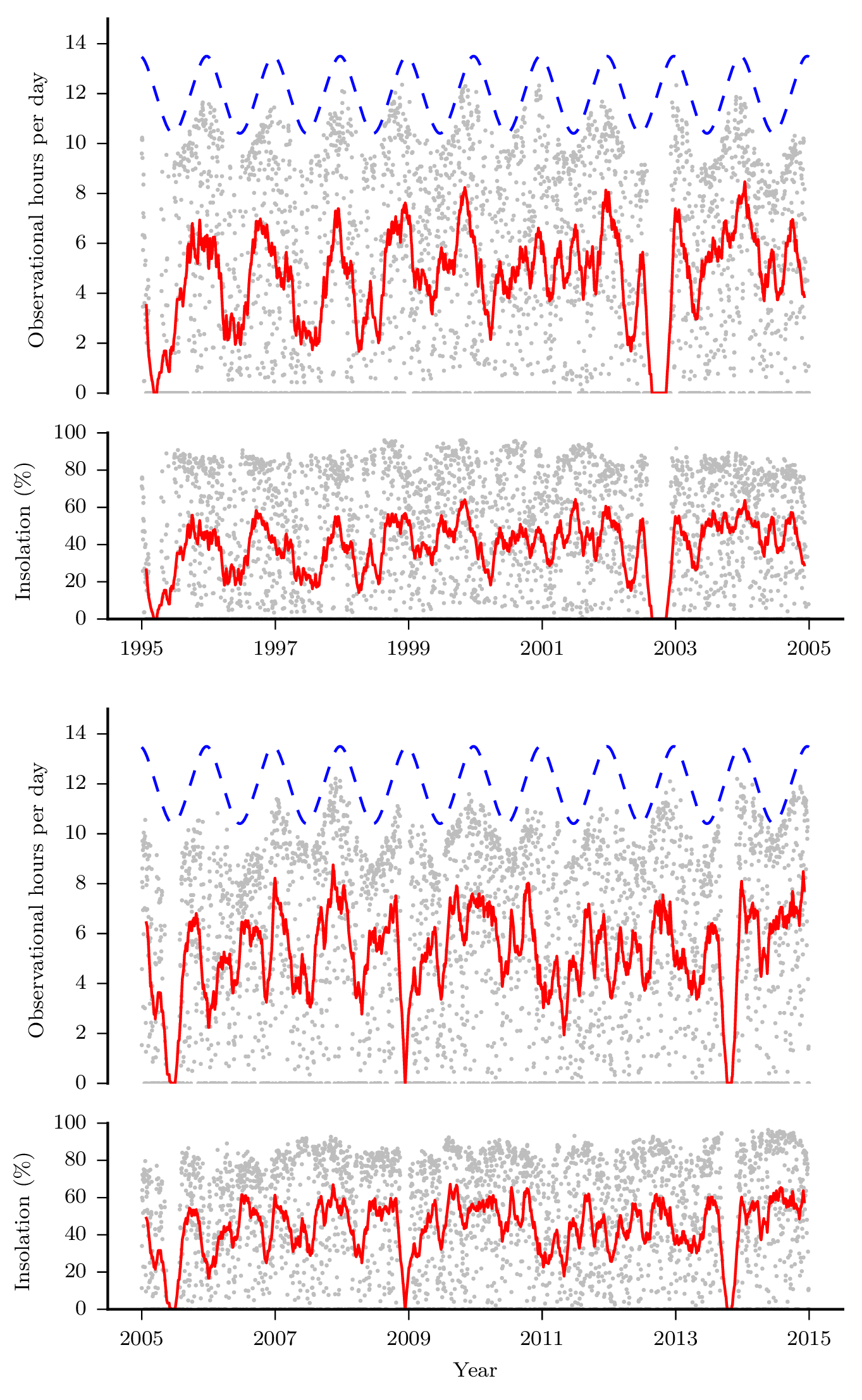}
\caption{Carnarvon duty cycle as a function of date, plotted in hours
  per day, and as a percentage of potential daylight hours. There is
  one grey dot per day, and the solid red curve represents a 50-day
  moving mean.  The dashed blue curve shows potential daylight hours.}
\label{fig:carnarvon_fill}
\end{figure}

\begin{figure}
\centering
\includegraphics[width=0.9\textwidth,height=0.9\textheight,keepaspectratio]{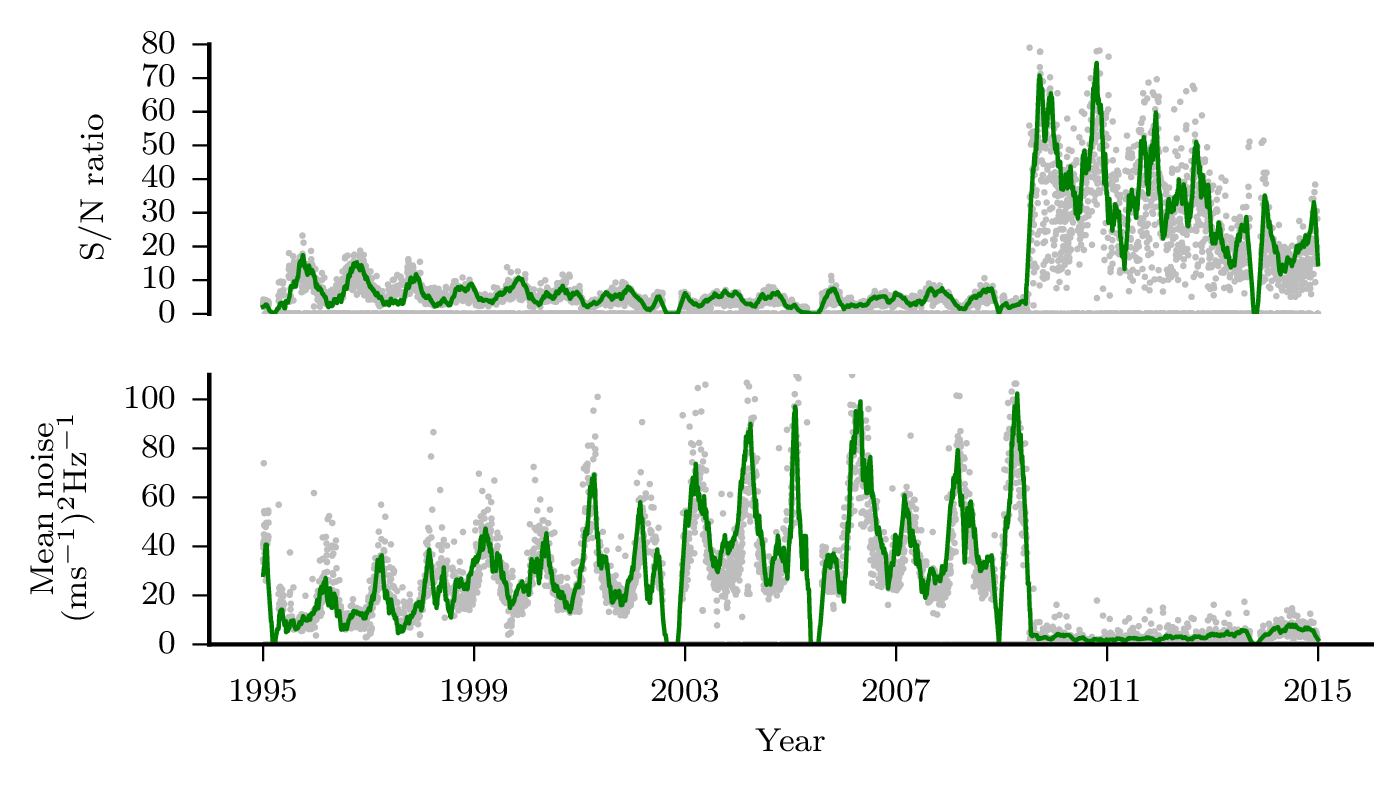}
\caption{Carnarvon data quality as a function of date.  Top: Signal to
  noise ratio, higher is better.  Bottom: Mean noise level, lower is
  better.  There is one grey dot per day, and the solid green curve
  represents a 50-day moving mean.  The large step in~2009 is due to
  installation of a new upgraded spectrometer.}
\label{fig:carnarvon_quality}
\end{figure}

Carnarvon has not been without its problems.  The PC failed in
August~2002, and this is the first drop-out in the Carnarvon duty
cycle (Figure~\ref{fig:carnarvon_fill}).  In May~2005 a freak
rainstorm emptied over three inches of rain in just two hours.  This
is almost the same amount of rain that Carnarvon expects in a whole
year.  It was not a good time to find out that the rain detector had
failed, and so the dome did not close~\citep{bison253,bison260}.  The
whole dome was thoroughly flooded, completely destroying the control
electronics for one instrument and severely damaging a second
instrument.  Water was poured out of some of the electronics.  At the
time, Carnarvon had two instruments in operation.  The primary
instrument was able to be repaired on-site after designing new
detectors and control electronics in Birmingham.  However, the
secondary instrument was written off and had to be returned to
Birmingham~\citep{bison282} for a considerable programme of repairs
and upgrades.  The new instrument was installed in~2009
\citep{bison323} and provided a substantial improvement in noise level
and data quality (Figure~\ref{fig:carnarvon_quality}).

In~2013 the land was bought by {NBN} Co.\ Ltd, the National Broadband
Network of Australia, as a base for a new satellite internet station.
A lease was approved between the group and NBN, and the dome was shut
down for just over a month in October~2013 during the NBN construction
in an attempt to limit the amount of dust and dirt entering the dome.
The NBN compound has two large antennas that are due east of the dome
and cause some shadowing in the early morning around the summer
solstice.  Despite this, the Carnarvon BiSON dome continues to collect
data year-round.

%%%%%%%%%%%%%%%%%%%%%%%%%%%%%%%%%%%%%%%%%%%%%%%%%%%%%%%%%%%%%%%%

\subsection{Sutherland, South Africa}

Following the success of the Carnarvon station, three more sites were
commissioned.  Some minor changes were made to the original design.
The main building became rectangular and made of brick, rather than
cylindrical and made from plywood clad with corrugated iron.  This
substantially increased the floor-area, allowing for more storage
space and easier access to the control electronics.  The mount was
also made considerably larger, allowing for bigger and heavier
instrumentation.

The first station of this new design was built in Birmingham in~1988.
It is located on the roof of the Poynting Physics building.  Once the
new-style dome had been finalised, the first site to be commissioned
was in Sutherland, South Africa, at the South African Astronomical
Observatory (SAAO) in~1989 \citep{bison226} and completed in~1990.
The observatory itself was established in~1972, and is run by the
National Research Foundation of South Africa.  It is located in the
Karoo Desert, in the Northern Cape of South Africa,
approximately~\SI{350}{\kilo\metre} northeast of Cape Town.

Like Carnarvon, Sutherland started out with a Keithley System~570
data-acquisition system, and later got its Zoo upgrade
in~2006 \citep{bison276}.  New digital temperature controllers were
installed in~2007 \citep{bison294,bison316}, new counters in~2012
 \citep{bison358}, and a new digital autoguider
in~2013 \citep{bison362}.

\begin{figure}
\centering
\includegraphics[width=0.9\textwidth,height=0.9\textheight,keepaspectratio]{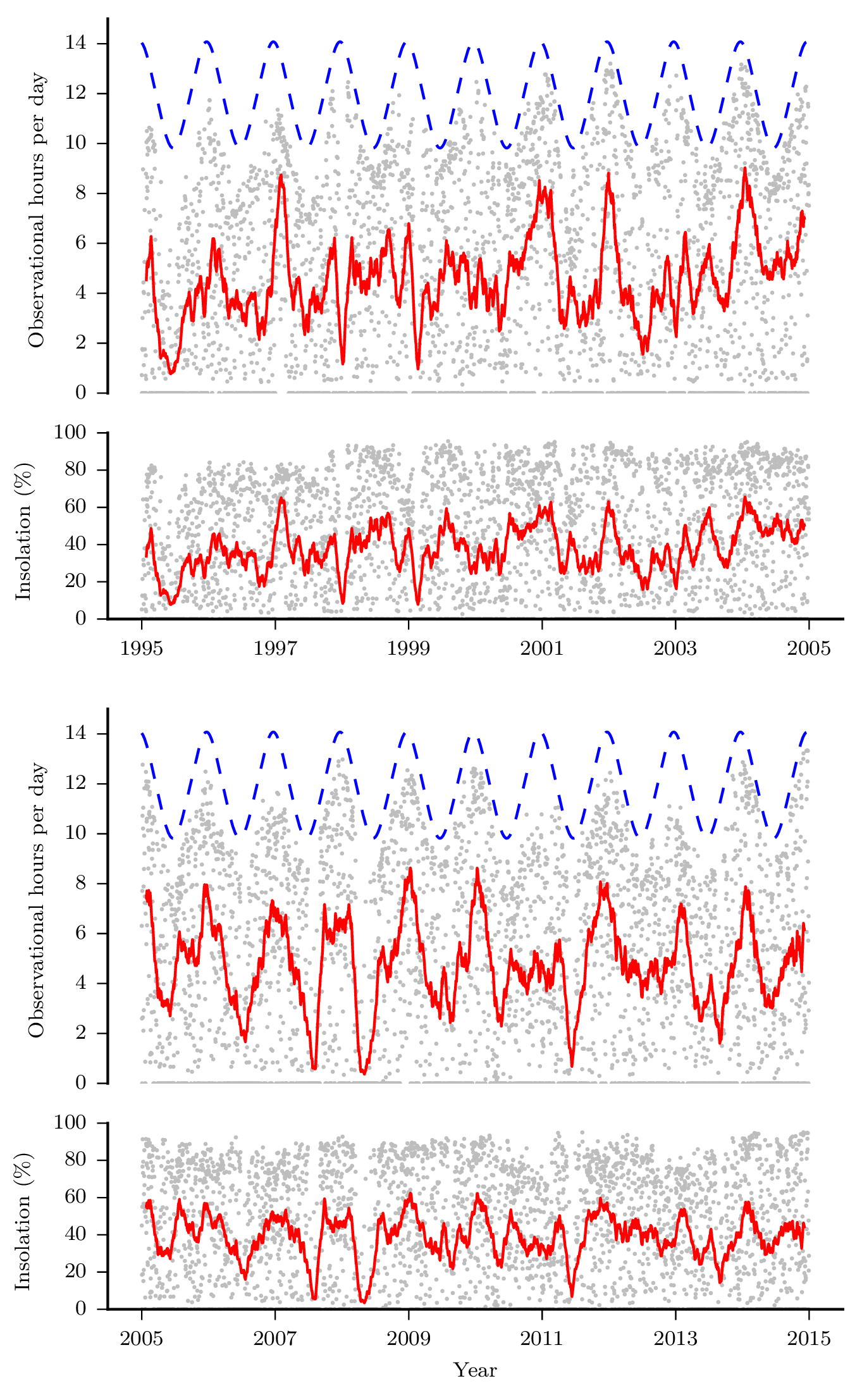}
\caption{Sutherland duty cycle as a function of date, plotted in hours
  per day, and as a percentage of potential daylight hours. There is
  one grey dot per day, and the solid red curve represents a 50-day
  moving mean.  The dashed blue curve shows potential daylight hours.}
\label{fig:sutherland_fill}
\end{figure}

The Sutherland duty cycle is shown in
Figure~\ref{fig:sutherland_fill}.  The drop-out in late~1997 through
early~1998 was caused by a range of faults.  The tracking motor on the
mount failed, and problems with the declination limit switches caused
the mount to be unresponsive.  Also, the dome moved out of alignment
and began shadowing the instrument just before sunset.  Later, the
electronic polarisation modulator failed, and this was not replaced
until late January~1998.  In early~1999, a fault developed on the
scaler system that counts the pulses produced by the detectors.
Following extensive investigation, simply re-seating all the chips on
the scaler cards fixed the problem, but not until almost a month of
data had been lost.  The problem reoccurred several times over the
years, until the scaler system was finally replaced completely
in~2012.  The gap in~2007 was due to some downtime whilst the
temperature control systems were upgraded, and in~2008 due to a heavy
snow storm.

\begin{figure}
\centering
\includegraphics[width=0.9\textwidth,height=0.9\textheight,keepaspectratio]{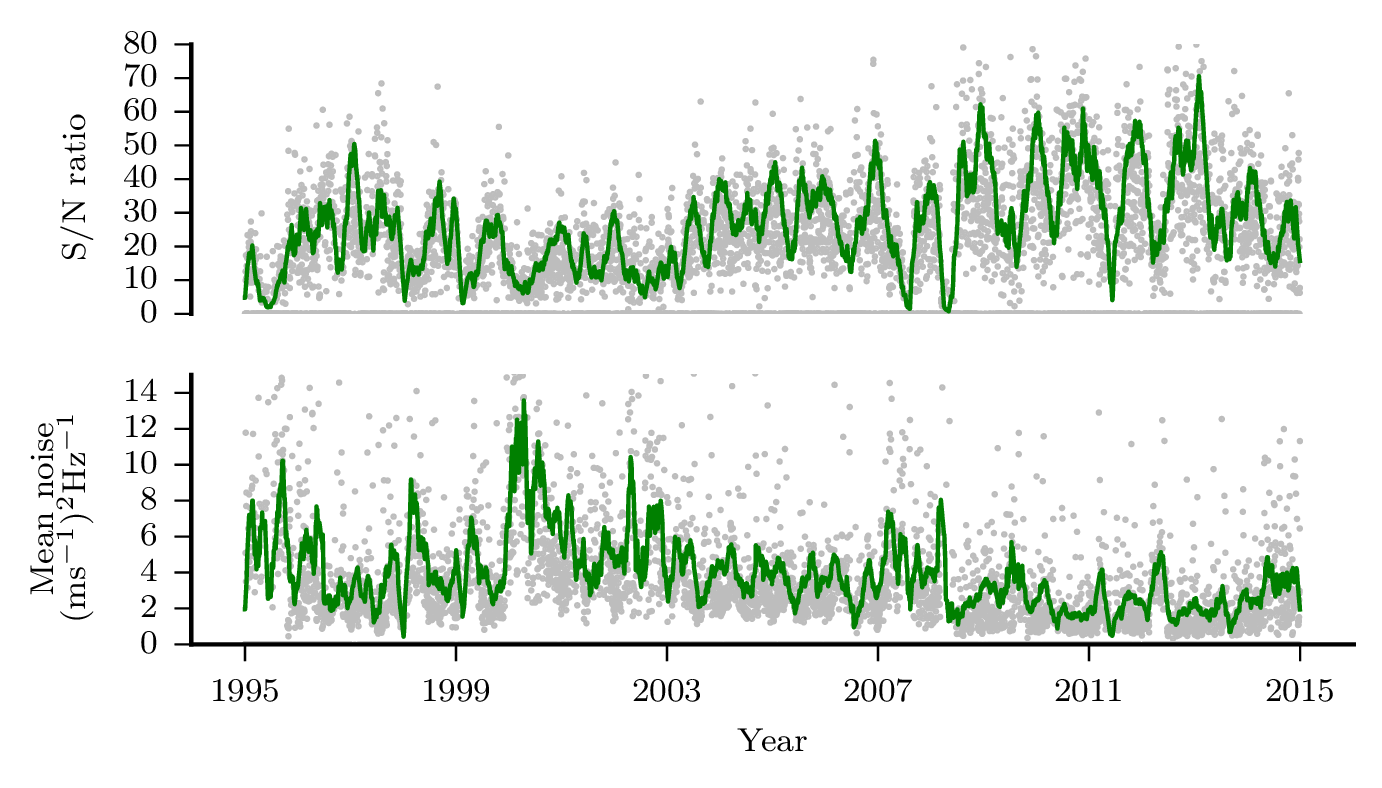}
\caption{Sutherland data quality as a function of date.  Top: Signal
  to noise ratio, higher is better.  Bottom: Mean noise level, lower
  is better.  There is one grey dot per day, and the solid green curve
  represents a 50-day moving mean.}
\label{fig:sutherland_quality}
\end{figure}

Looking at the data quality and noise levels
(Figure~\ref{fig:sutherland_quality}) we see that the site has shown
consistent performance with gradual improvement as systems were
upgraded.  The higher noise level in~2000 was due to a sticky
declination gearbox.  Weather conditions in Sutherland show solid
year-round performance.

%%%%%%%%%%%%%%%%%%%%%%%%%%%%%%%%%%%%%%%%%%%%%%%%%%%%%%%%%%%%%%%%

\subsection{Las Campanas, Chile}

After the completion of Sutherland, a third automated site was opened
in Chile in~1991, at the Las Campanas Observatory operated by the
Carnegie Institution for Science.  Las Campanas Observatory is located
in the southern Atacama Desert of Chile, around~\SI{100}{\kilo\metre}
northeast of La Serena.  The observatory was established in~1969, and
was a replacement for the Mount Wilson Hale Observatory near Pasadena
which had started to experience too much light pollution from the
growing city of Los Angeles.  The main office is in Las Serena, whilst
the headquarters remain in Pasadena.

Las Campanas received its Zoo upgrade in
December~2005 \citep{bison261}, a new digital autoguider in
February~2011 \citep{bison343,bison344}, and a new temperature
controller in~2015 \citep{bison372,bison373}.

\begin{figure}
\centering
\includegraphics[width=0.9\textwidth,height=0.9\textheight,keepaspectratio]{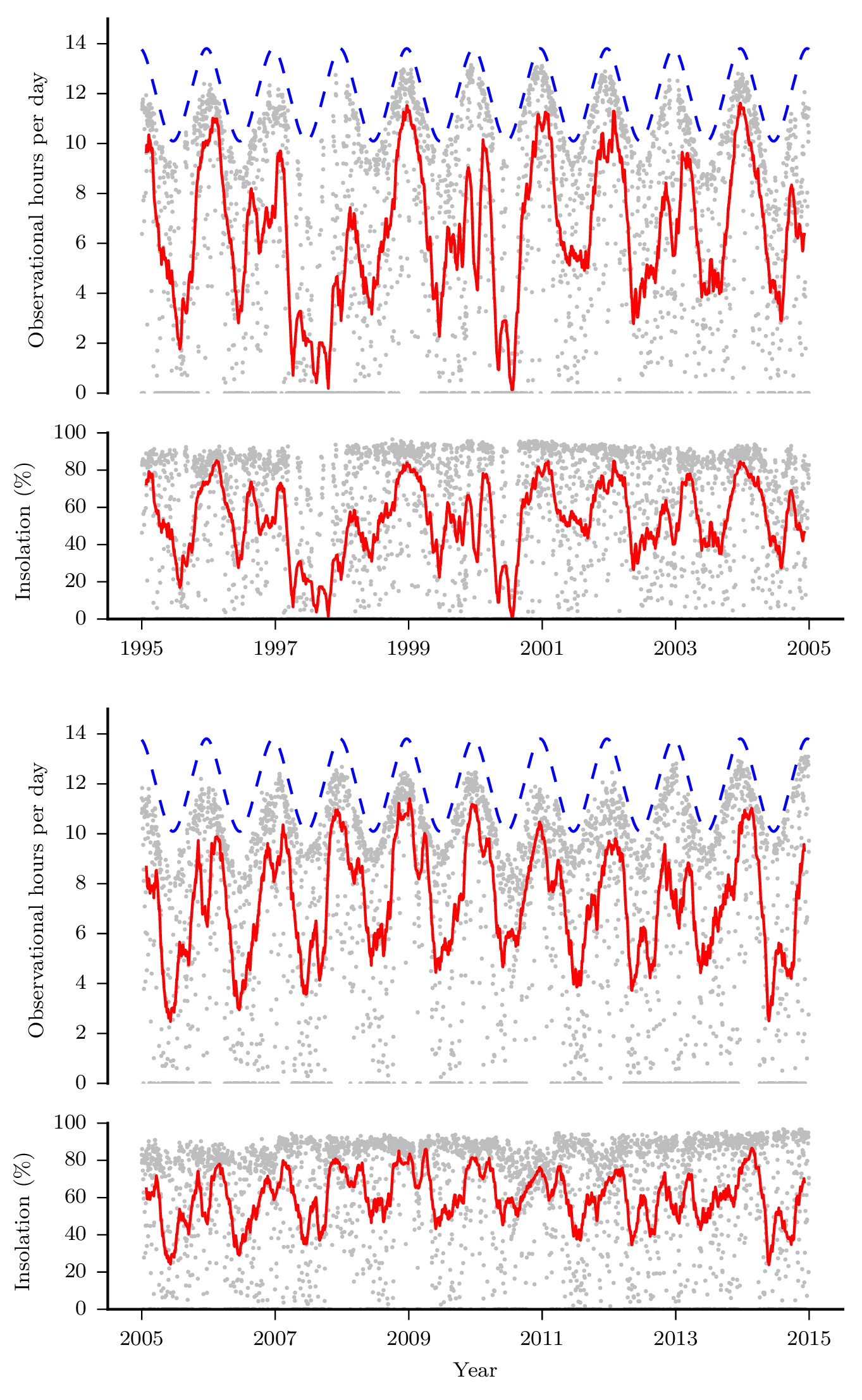}
\caption{Las Campanas duty cycle as a function of date, plotted in
  hours per day, and as a percentage of potential daylight
  hours. There is one grey dot per day, and the solid red curve
  represents a 50-day moving mean.  The dashed blue curve shows
  potential daylight hours.}
\label{fig:campanas_fill}
\end{figure}

The Las Campanas duty cycle is shown in
Figure~\ref{fig:campanas_fill}.  Las Campanas has had a range of
problems.  One notable experience was in July~1997 when a direct
lightning strike to the dome destroyed several pieces of sensitive
electronics.  Whilst repairs were completed, weather conditions
deteriorated with a heavy snow fall that left the observatory cut off
by road, without internet or phone connections, and dependent on old
diesel generators for electrical power~\citep{bison062}.  As the snow
melted, the~\SI{4.2}{\kilo\volt} step-down transformer for the {BiSON}
dome was first flooded, and subsequently destroyed by another
lightning strike.  The dome was without power for more than a month,
until early November~1997.

\begin{figure}
\centering
\includegraphics[width=0.9\textwidth,height=0.9\textheight,keepaspectratio]{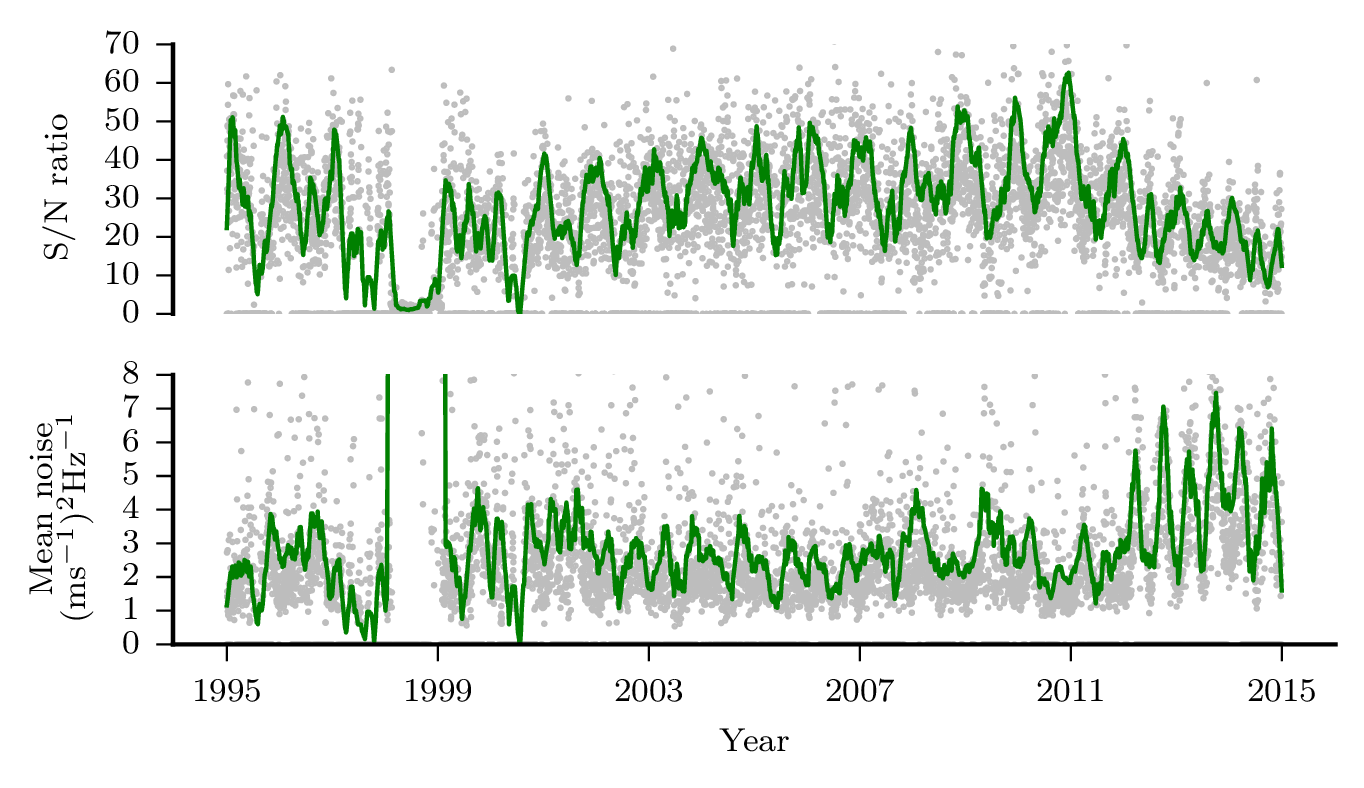}
\caption{Las Campanas data quality as a function of date.  Top: Signal
  to noise ratio, higher is better.  Bottom: Mean noise level, lower
  is better.  There is one grey dot per day, and the solid green curve
  represents a 50-day moving mean.}
\label{fig:campanas_quality}
\end{figure}

Two further visits to finalise repairs were needed in November~1997
\citep{bison071} and January~1998 \citep{bison074}, which were related
to problems with the computer, the dome azimuth motor, and the
water-loop system that stabilises the instrumentation temperatures.
Noise problems in both detectors persisted until early~1999, clearly
visible in the plot of data quality
(Figure~\ref{fig:campanas_quality}).  A large part of~2000 was lost
due to a broken declination gearbox on the mount, and also to the
failure of the dome azimuth motor.  More electrical problems occurred
in May~2014 when faults with both the shutter and blind limit switches
caused the circuit breakers to trip repeatedly.  A site visit was
required to replace the limit switches, and also work on additional
problems with the water-loop pump and the uninterruptable power
supplies~\citep{bison366}.

Las Campanas is the best performing station in the network,
consistently supplying duty cycles above 80\,\% in the summer and
regularly above 40\,\% even during the winter months.  From~2012 the
noise performance has deteriorated slightly.  A recent site visit
indicated reduced performance from the potassium vapour cell, and this
may need to be replaced soon.

%%%%%%%%%%%%%%%%%%%%%%%%%%%%%%%%%%%%%%%%%%%%%%%%%%%%%%%%%%%%%%%%

\subsection{Narrabri, NSW, Australia}

The final fully-automated site was installed in Narrabri, Australia,
in~1992 \citep{bison227}.  It is on the site of the Australia
Telescope Compact Array (ATCA) at the Paul Wild Observatory operated
by The Commonwealth Scientific and Industrial Research Organisation
(CSIRO).  The observatory is around~\SI{550}{\kilo\metre} northwest of
Sydney.

Narrabri received its Zoo upgrade in June~2004 \citep{bison241}, new
temperature controllers in January~2010 \citep{bison332,bison333}, and
a new digital autoguider and counters in April~2013 \citep{bison360}.
In February~2000 the dome blind motor failed \citep{bison138} and was
replaced with an identical spare from Birmingham, but failed again in
March~2003 by which time the part had been discontinued.  The dome
manufacturer specified an alternative part for the blind mechanism,
and this was installed in July~2003 \citep{bison213}.  Following
several futher motor failures a change to the control system was
indentified as being required due differences in the motor
type~\citep{bison241}.

\begin{figure}
\centering
\includegraphics[width=0.9\textwidth,height=0.9\textheight,keepaspectratio]{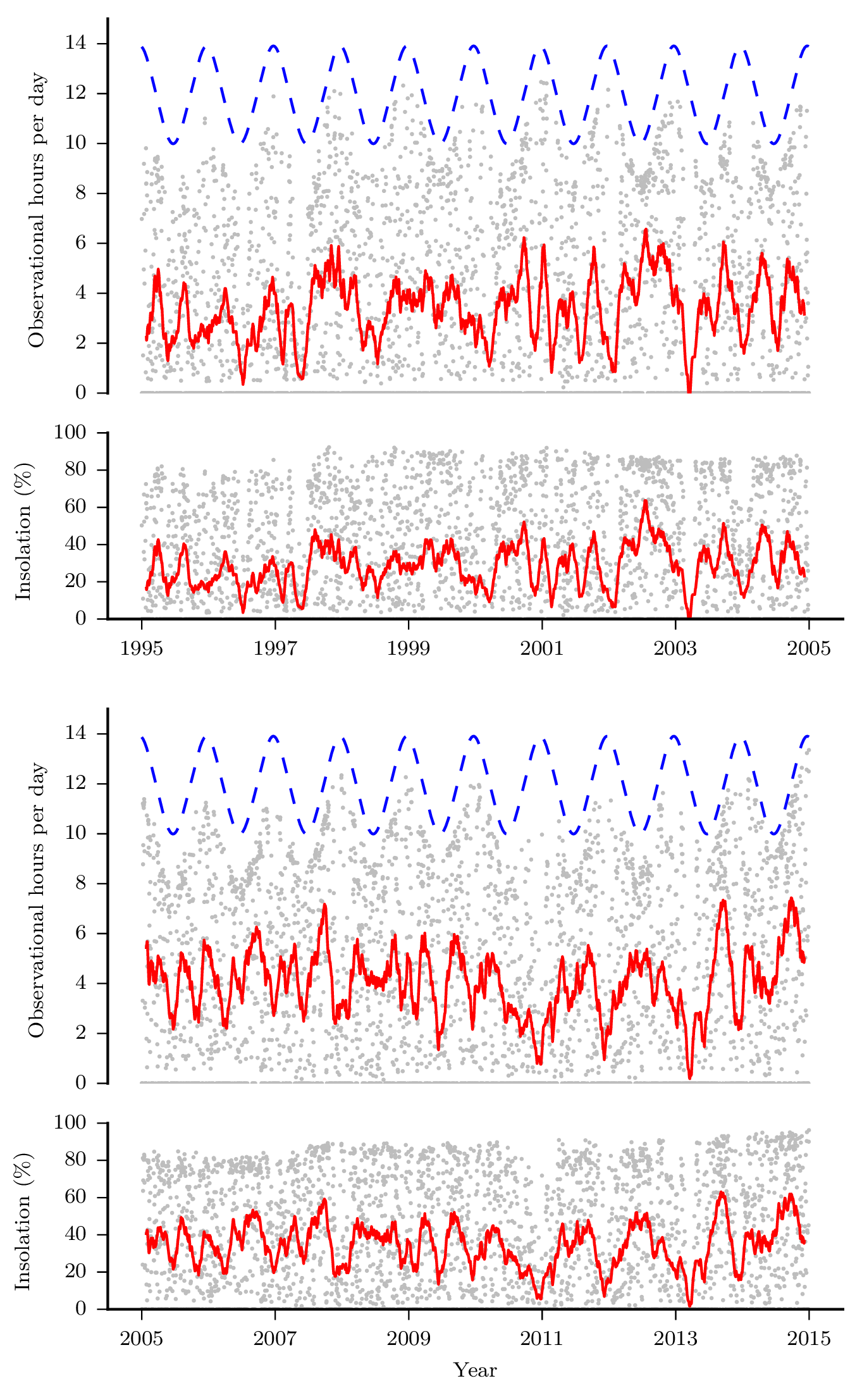}
\caption{Narrbri duty cycle as a function of date, plotted in hours
  per day, and as a percentage of potential daylight hours. There is
  one grey dot per day, and the solid red curve represents a 50-day
  moving mean.  The dashed blue curve shows potential daylight hours.}
\label{fig:narrabri_fill}
\end{figure}

The Narrabri duty cycle is shown in Figure~\ref{fig:narrabri_fill}.
The gap in~1997 is due to problems with the scalers and correct
termination of the signal cables from the voltage-to-frequency output
of the detectors.  The missing data in~2003 and~2004 are due to the
blind motor problems discussed earlier.  In~2013 a fault on the
anemometer in June caused problems with the weather module, and this
kept the dome closed unnecessarily.

\begin{figure}
\centering
\includegraphics[width=0.9\textwidth,height=0.9\textheight,keepaspectratio]{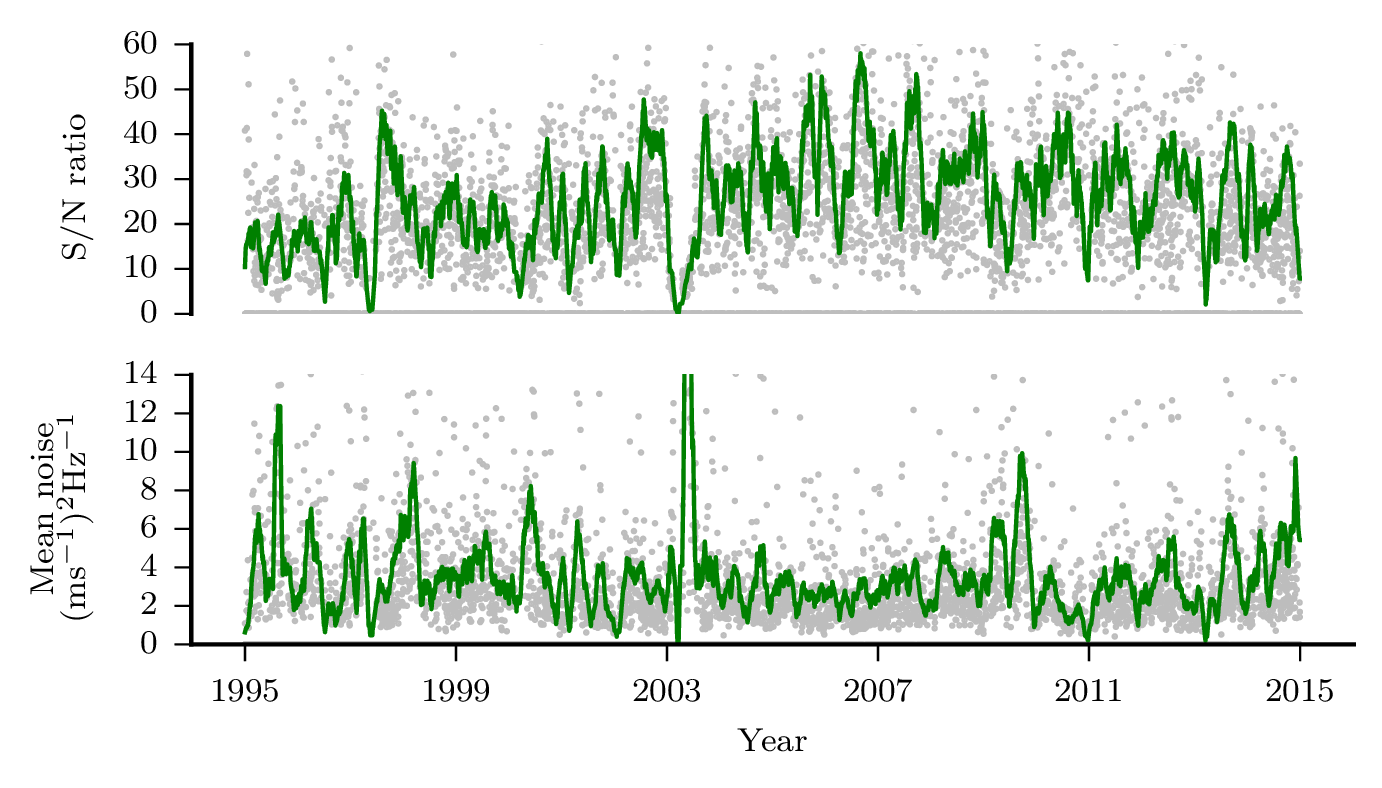}
\caption{Narrabri data quality as a function of date.  Top: Signal to
  noise ratio, higher is better.  Bottom: Mean noise level, lower is
  better.  There is one grey dot per day, and the solid green curve
  represents a 50-day moving mean.}
\label{fig:narrabri_quality}
\end{figure}

In the plot of data quality, Figure~\ref{fig:narrabri_quality}, the
increased noise level in~2003 is due to a failure of an interference
filter temperature controller, and in~2009 is due to a faulty power
supply producing under-voltage power rails.  Other than these faults,
the site shows solid performance.

%%%%%%%%%%%%%%%%%%%%%%%%%%%%%%%%%%%%%%%%%%%%%%%%%%%%%%%%%%%%%%%%

\subsection{Mount Wilson, California, USA}

The final change to arrive at the existing network configuration was
made in~1992.  The {Mark~III} instrument was moved from Hawaii to the
60~foot tower at the Mount Wilson Hale Observatory in
California~\citep{bison005}.  This is the very same observatory where
Robert Leighton originally discovered solar five-minute oscillations
\citep{1962ApJ...135..474L}. The observatory is located in the San
Gabriel Mountains near Pasadena, around~\SI{80}{\kilo\metre} northeast
of Los Angeles.  The tower is operated by Ed Rhodes and his team of
undergraduate volunteer observers.  Like Iza\~na the tower uses a
c{\oe}lostat to direct light down the tower into the observing room
below, and as such requires someone to open the dome and align the
mirrors each morning.  It also requires presence throughout the day to
close the dome in the event of bad weather, and at the end of the
observing session.

The original optical configuration used by Leighton involved the
c{\oe}lostat firing light 60~feet down the tower, a further 30~feet
down into a pit where it would reflect from the spectroheliograph, and
finally another 30~feet back up the pit to the observing room above,
the total optical path length being 120~feet.  The spectroheliograph
is no longer in use, and so the current optical configuration is
somewhat different.  The primary instrument operated by Rhodes used a
small objective lens near the base of the tower.  Since it required
only half the original optical path length, the c{\oe}lostat mirrors
are now effectively oversized for their current usage.  We can take
advantage of this by ``picking off'' a small section of the beam and
directing it to another instrument without affecting the operation of
the main instrument.  This is done using a ``periscope'' arrangement
of mirrors mounted in the tower shaft taking a part of the beam and
shifting it slightly to the south where it is directed back down the
tower.  Another mirror at the bottom reflects the light horizontally
into the {BiSON} spectrometer.  In total, five mirrors are used to
direct light into the spectrometer.

\begin{figure}
\centering
\includegraphics[width=0.9\textwidth,height=0.9\textheight,keepaspectratio]{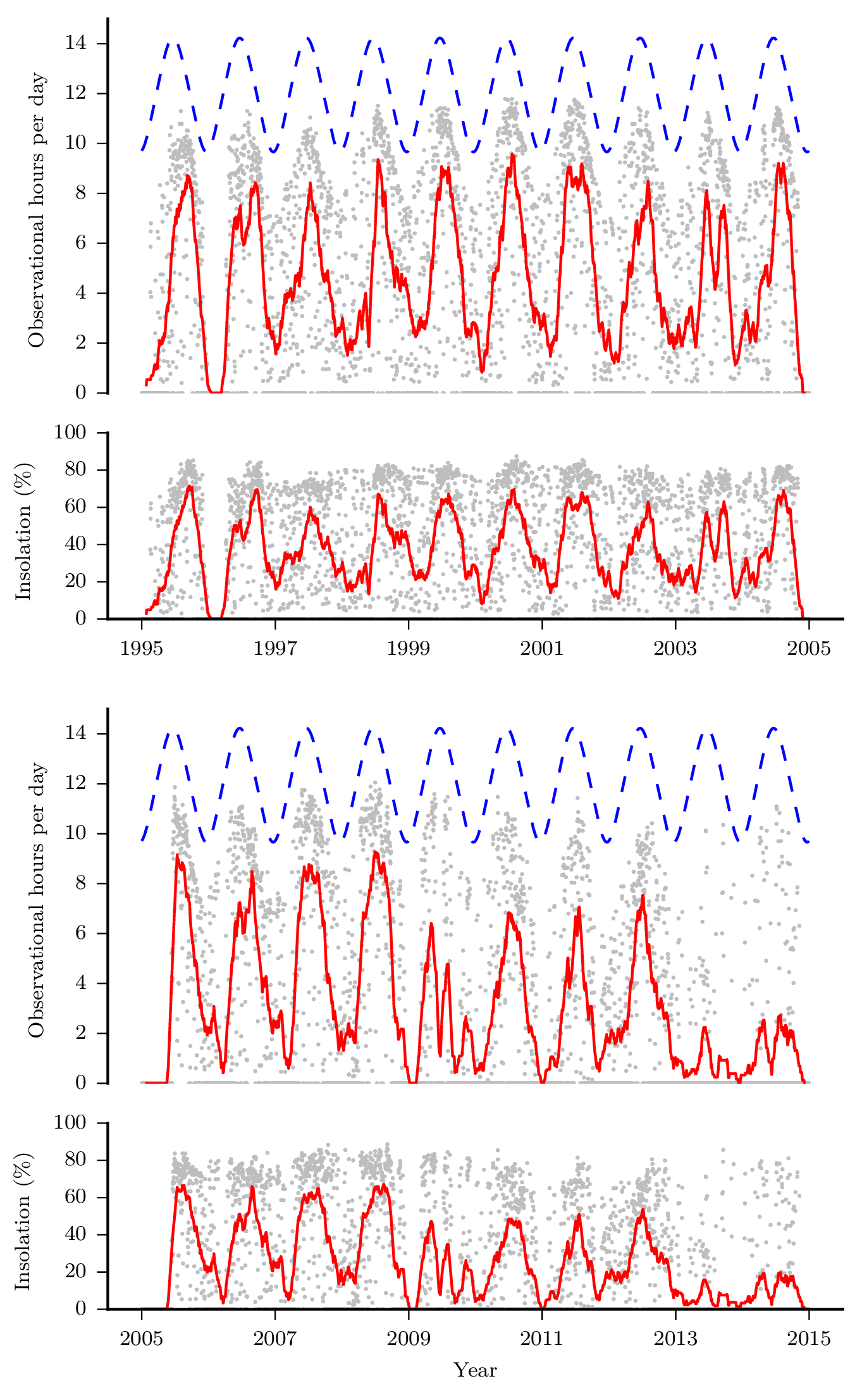}
\caption{Mount Wilson duty cycle as a function of date, plotted in
  hours per day, and as a percentage of potential daylight
  hours. There is one grey dot per day, and the solid red curve
  represents a 50-day moving mean.  The dashed blue curve shows
  potential daylight hours.}
\label{fig:mountwilson_fill}
\end{figure}

There are regular gaps in the data at midday during the winter months,
when the primary mirror has to be moved from the east side of the
tower in the morning to the west in the afternoon, in order to avoid
shadowing from the secondary mirror.  Unlike Iza\~na, the secondary at
Mount Wilson has only one mounting configuration.  The Mount Wilson
duty cycle is shown in Figure~\ref{fig:mountwilson_fill}.

\begin{figure}
\centering
\includegraphics[width=0.9\textwidth,height=0.9\textheight,keepaspectratio]{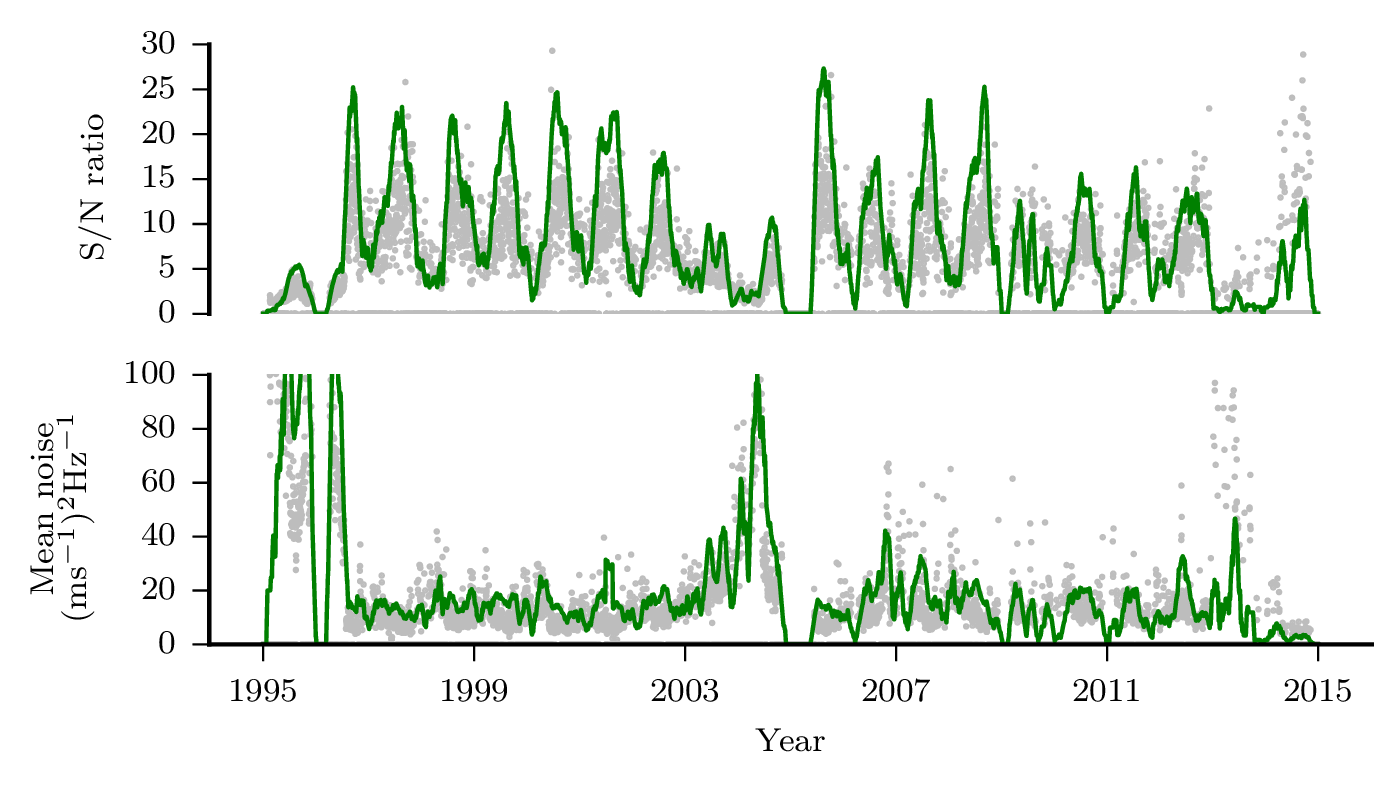}
\caption{Mount Wilson data quality as a function of date.  Top: Signal
  to noise ratio, higher is better.  Bottom: Mean noise level, lower
  is better.  There is one grey dot per day, and the solid green curve
  represents a 50-day moving mean.  The large step in~1996 is due to
  installation of a new upgraded spectrometer, and in~2004 due to
  autoguider problems.}
\label{fig:mountwilson_quality}
\end{figure}

In July~1996 the {Mark~III} instrument from Haleakala was retired and
replaced with the tenth spectrometer designed by the group, code-named
Klaus~\citep{bison106}.  The new design offered much improved data
quality in line with the other newer instruments in the network, and
the reduction in noise level can be seen clearly in the plot of data
quality (Figure~\ref{fig:mountwilson_quality}).

The optical configuration with the pick-off mirrors is marginal, and
very precise alignment is required for optimum performance of the
spectrometer.  Aligning five different mirrors over such large
distances and ensuring no vignetting is difficult.  Vibrations from
the tower can also cause the mirrors or the spectrometer itself to
move slightly, meaning the optical alignment has to be checked
frequently.  The large step in noise level during~2004 corresponds to
an autoguider problem, where it failed to drive in right ascension.
Mount Wilson received its Zoo upgrade in September~2005
\citep{bison255}, which included alignment of all the optics returning
the site to its original performance.  In~2009 a variety of faults
caused loss of data.  The computer failed at the beginning of the
year, and a site visit to replace it could not be arranged until
March.  Shortly after installing the new computer, the UPS failed in
May and caused problems with the temperature controller.  The on-site
help investigated the problem but could not locate the fault.  Another
site visit was arranged for July, where a blown fuse was found and the
problem solved.  At the end of July the mirrors were removed for two
weeks for re-aluminisation.  Just at the point where everything was
working, in August the entire observatory was evacuated due to the
nearby ``Station Fire''.  There was concern that the entire
observatory could be lost.  Fortunately that was not the case, thanks
to efforts from the fire service and the US Forestry Service.

Further guider problems were experienced in~2011, where one of the
motors failed again and needed to be replaced.  Through~2013 and~2014,
significant problems were experienced with the autoguider electronics
in the tower.  On a site visit from Birmingham, the guider system
built into the second flat of the c{\oe}lostat was completely rewired
and refurbished, removing some very old and corroded cables and
producing a significant decrease in noise level~\citep{bison365}.
Significant damage to the primary mirror was also discovered during
the visit.  This is a serious problem since it is implicit in our
analysis that we see the whole of the solar disc and that no part is
vignetted.  Since the Sun is rotating, any vignetting of the disc
causes uneven weighting and produces an offset in the computed
residuals through a process known as Doppler
Imaging~\citep{1978MNRAS.185...19B}.  For the Sun, when viewing a
typical Fraunhofer line the ratio in weighting of opposite sides of
the disc is about 4:1.  The instrument used at Mount Wilson is the
same design as used at our other sites on equatorial mounts.  It is
optically designed to directly observe the Sun.  At Mount Wilson, the
c{\oe}lostat is approximately 60~feet away from the spectrometer, at
the top of the tower.  Optically this is near infinity and so the
spectrometer forms an image of the surface of the c{\oe}lostat
mirrors, with the result that it is essential the mirrors be clean and
in good condition.  The damaged area of the mirror was causing part of
the solar disc to be missing from the light entering the spectrometer.
The problem was made worse due to the unusual design of the tower
c{\oe}lostat.  Usually, the primary mirror performs both tracking and
guiding to follow the Sun throughout the day.  The Mount Wilson
c{\oe}lostat separates these functions between the two mirrors -- the
primary mirror tracks whilst the secondary mirror guides.  This means
that any dirt or damage on the surface of the primary mirror will give
rise to a signal that will be seen to oscillate at the frequency of
the tracking error, as the autoguider on the secondary mirror
compensates for tracking errors of the primary mirror.  The worm-drive
of the primary mirror uses 432~teeth, which puts the expected ``gear
frequency'' at~\SI{2.5}{\milli\hertz}.  This causes the vignetted part
of the solar image to oscillate at~\SI{2.5}{\milli\hertz},
periodically varying the weighting of the solar disc, and corrupting
the power spectrum at this frequency with sufficient amplitude to make
the data unusable.  Unfortunately it is almost impossible to filter
out this fault since it is within the main solar five-minute
oscillation band of interest, falling very close to one of the mode
peaks, and is not sufficiently coherent to be removed by subtracting a
single sine wave.  The fault was able to be mitigated by rotating the
mirror in its housing to move most of the damage to an area not used
by our pick-off mirrors.  A similar fault occurs if the two mirrors
are not aligned correctly at the start of an observing period.  If the
light from the primary mirror ``falls off'' the edge of the secondary
mirror, then again the solar disc is vignetted and
the~\SI{2.5}{\milli\hertz} tracking error becomes visible.

%%%%%%%%%%%%%%%%%%%%%%%%%%%%%%%%%%%%%%%%%%%%%%%%%%%%%%%%%%%%%%%%

%% file: network.tex
% NETWORK.TEX
%
%   Steven Hale
%   2015 June 1
%   Birmingham, UK
%
% BiSON Statistics - Whole Network
%
% $Id: network.tex 59 2015-10-17 06:44:36Z hale $
%

%%%%%%%%%%%%%%%%%%%%%%%%%%%%%%%%%%%%%%%%%%%%%%%%%%%%%%%%%%%%%%%%

\section{Whole Network Performance}
     \label{S-Whole Network Performance}

%%%%%%%%%%%%%%%%%%%%%%%%%%%%%%%%%%%%%%%%%%%%%%%%%%%%%%%%%%%%%%%%

\begin{figure}
\centering
\includegraphics[width=0.9\textwidth,height=0.9\textheight,keepaspectratio]{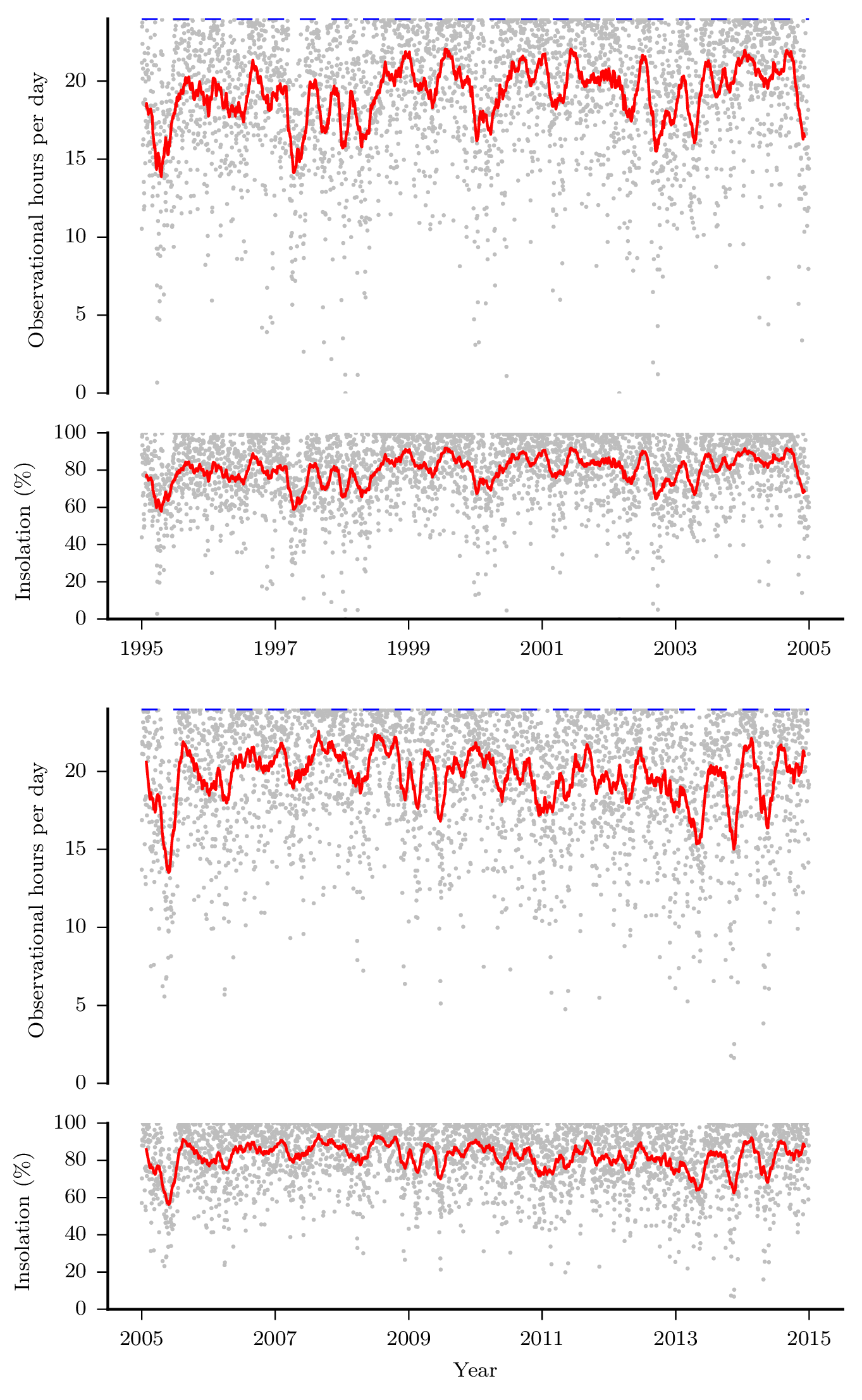}
\caption{All-station duty cycle as a function of date, plotted in
  hours per day, and as a percentage of potential daylight
  hours. There is one grey dot per day, and the red curve represents a
  50-day moving mean.}
\label{fig:all_fill}
\end{figure}

\begin{figure}
\centering
\includegraphics[width=0.9\textwidth,height=0.9\textheight,keepaspectratio]{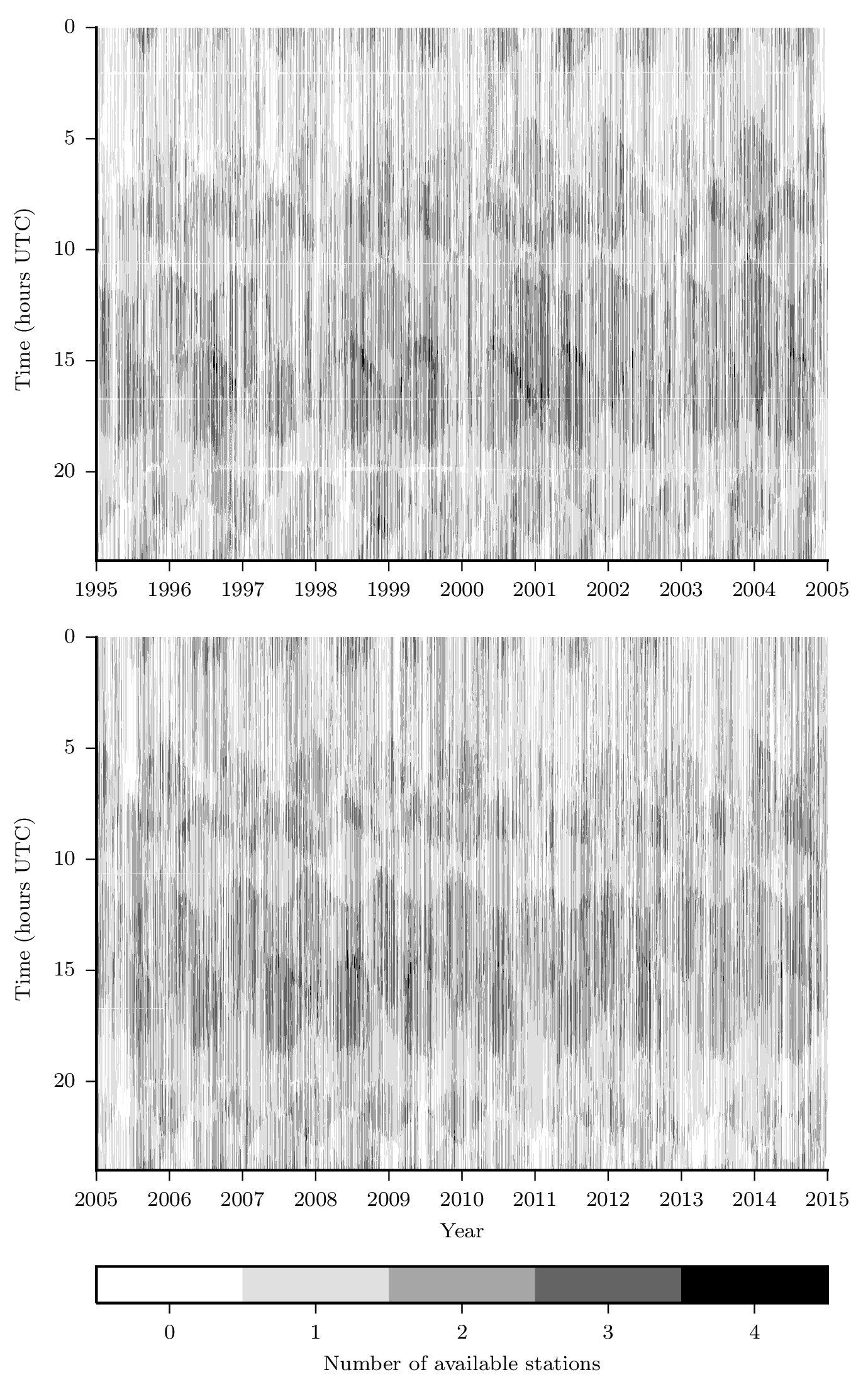}
\caption{All-station data window-function.}
\label{fig:all_window}
\end{figure}

\begin{figure}
\centering
\includegraphics[width=0.9\textwidth,height=0.9\textheight,keepaspectratio]{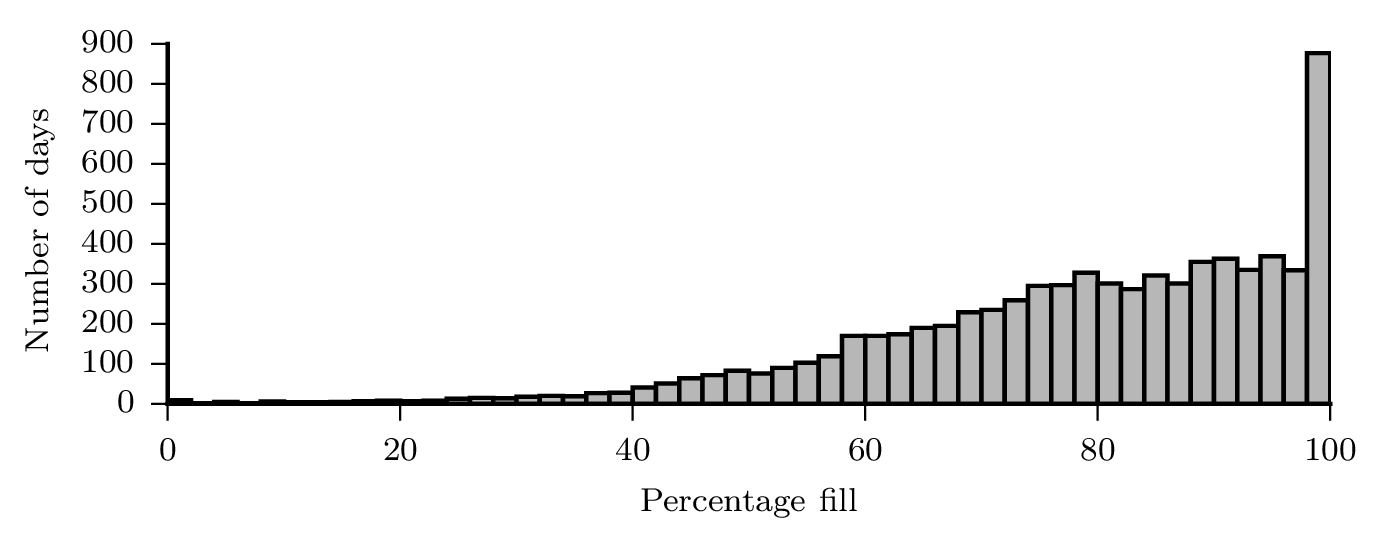}
\caption{The distribution of fill per day for the overall time series.}
\label{fig:all_fillhist}
\end{figure}

Plots of the whole-network duty cycle and data window-function are
shown in Figures~\ref{fig:all_fill} and~\ref{fig:all_window}.  In a
total of 7305 days, 45.5\,\% of the available time was covered by one
site, 30.3\,\% by two sites, 6\,\% by three sites, 0.06\,\% by four
simultaneous sites.  The horizontal lines that run from 1995\,--\,2005
in Figure~\ref{fig:all_window} are caused by a midday beam-chopper
used to check the dark counts from the detectors during the day.
From~2006 it was decided that this was no longer necessary, and so the
regular gap is not present.  Other regular gaps are those caused by
c{\oe}lostat shadowing at Iza\~na and Mount Wilson, as discussed in
Section~\ref{S-Site Performance}.  Figure~\ref{fig:all_fillhist} shows
a histogram of the daily fill.  Just 9~days had no coverage, and
633~days have a fill greater than 99\,\% of which 312~days achieved
100\,\%.  The average fill is 82\,\% for the whole dataset.

The original data pipeline for calibration of raw data from the
{BiSON} spectrometers through to velocity residuals is described by
\cite{1995A&AS..113..379E}.  The next stage of analysis involves
combining the residuals into an extended time series, and transforming
into the frequency domain where the mode characteristics can be
analysed.  This is described by \cite{1997A&AS..125..195C} and
\cite{halemphil}.  An updated pipeline that includes correction for
differential atmospheric extinction was produced by
\cite{2014MNRAS.441.3009D}.  By applying differential extinction
correction we have removed most of the low-frequency drifts in the
dataset that were previously filtered using a 25-sample moving mean,
and this allows further investigation of the very low frequency modes
of oscillation that were previously lost in the noise background.  We
are now also able to make better use of weighted averaging of
overlapping sites to produce a further improvement in signal to noise
ratio.
    
\begin{figure}
\centering
\includegraphics[width=0.9\textwidth,height=0.9\textheight,keepaspectratio]{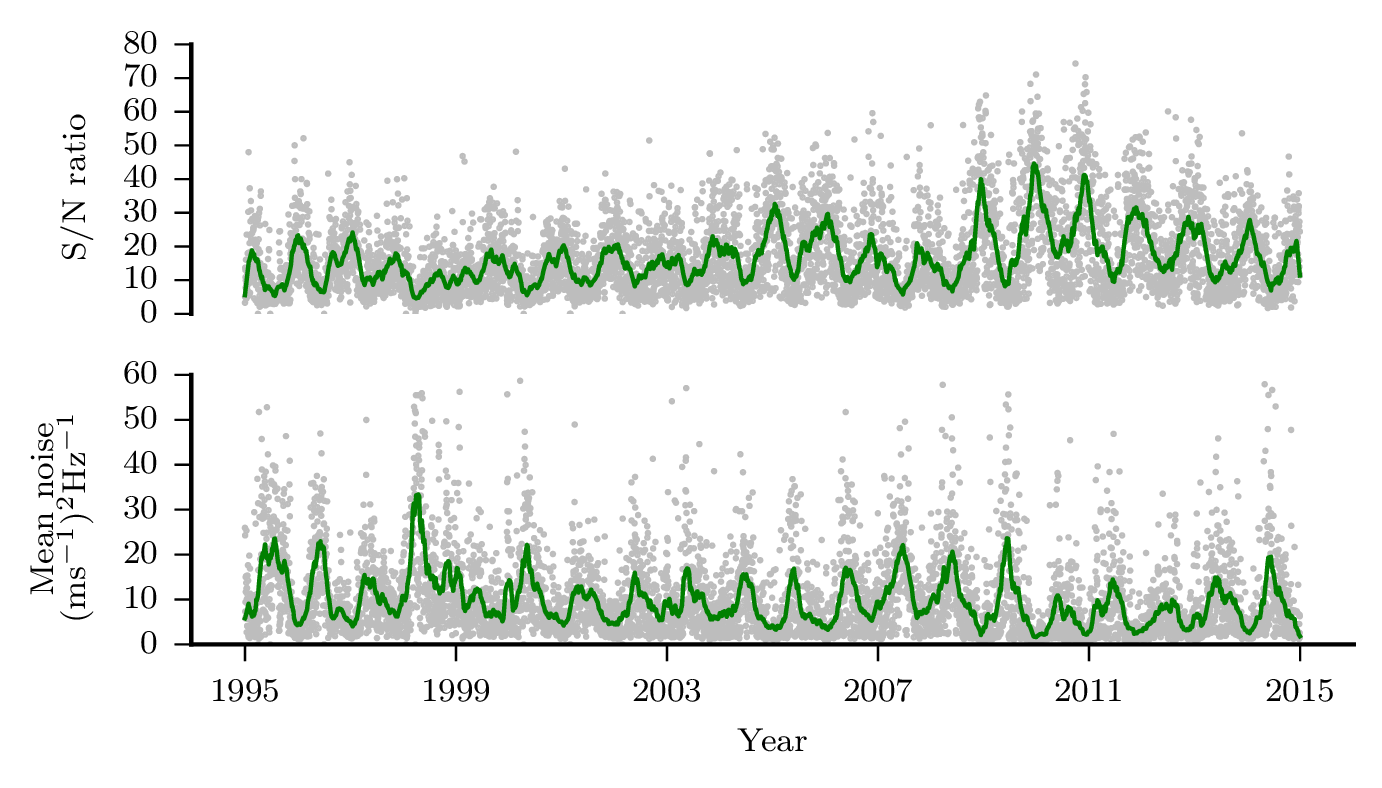}
\caption{All station data quality as a function of date.  Top: Signal
  to noise ratio, higher is better.  Bottom: Mean noise level, lower
  is better.  There is one grey dot per day, and the solid green curve
  represents a 50-day moving mean.}
\label{fig:all_quality}
\end{figure}

\begin{figure}
\centering
\includegraphics[width=0.9\textwidth,height=0.9\textheight,keepaspectratio]{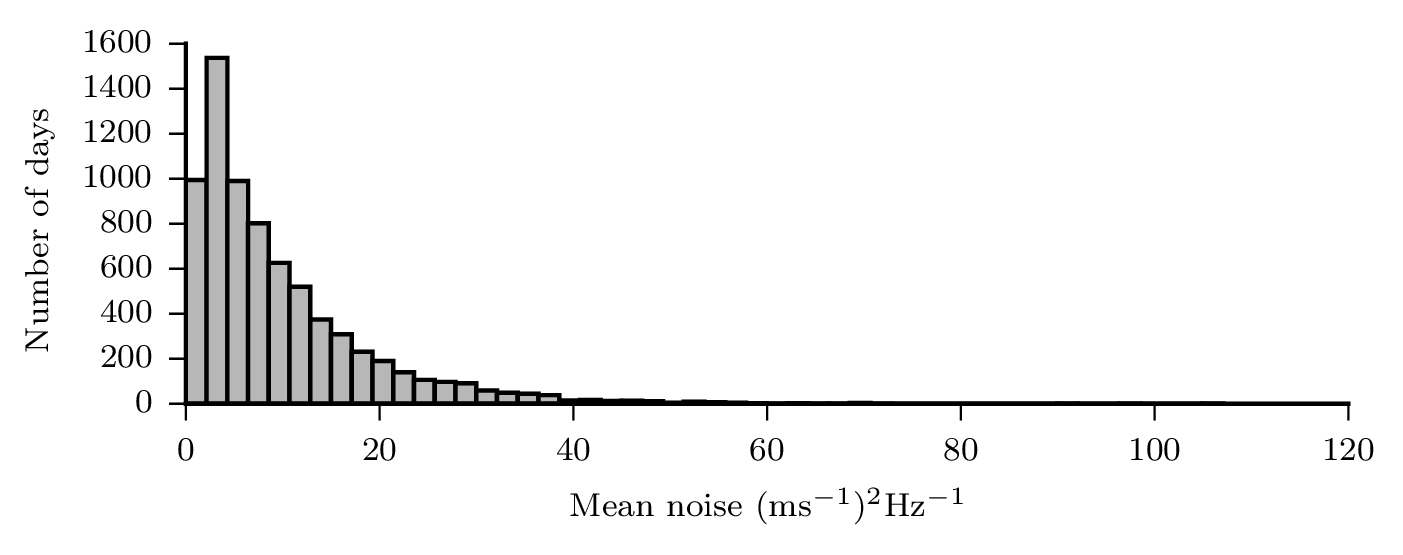}
\caption{The distribution of mean noise per day for the overall time
  series.  Lower is better.}
\label{fig:all_noisehist}
\end{figure}

\begin{figure}
\centering
\includegraphics[width=0.9\textwidth,height=0.9\textheight,keepaspectratio]{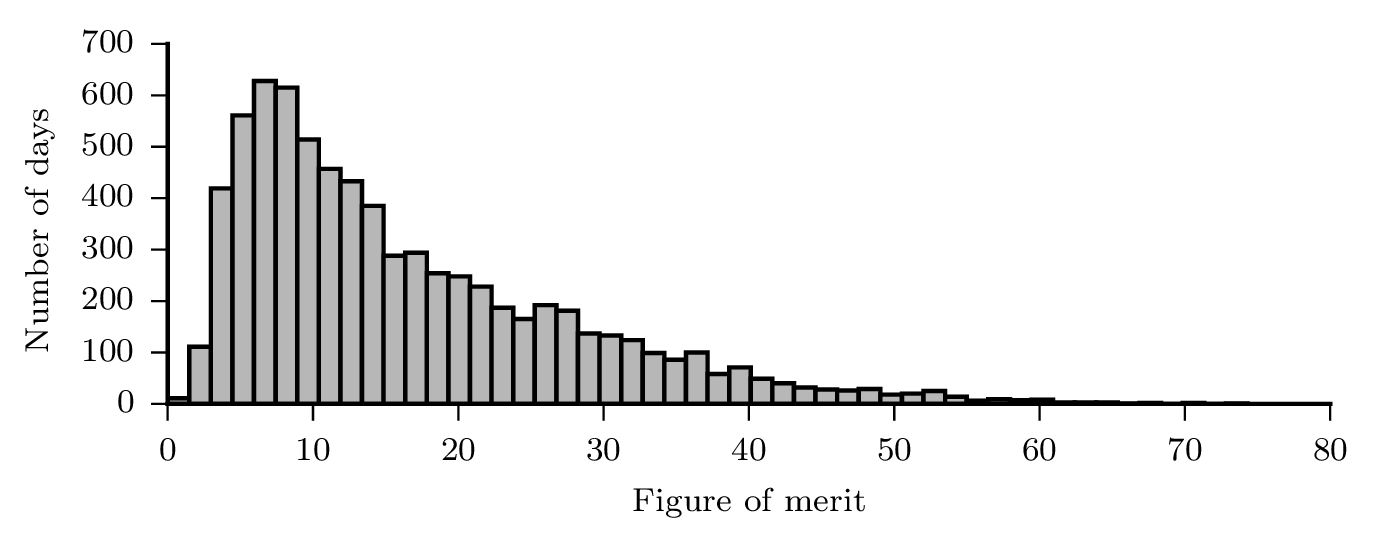}
\caption{The distribution of FOM per day for the overall time
  series.  Higher is better.}
\label{fig:all_fomhist}
\end{figure}

The entire network archive of velocity residuals has been regenerated
using the latest data pipeline, and a new concatenated time series has
been produced for the period from 1995 to the end of 2014 using the
latest weighted-averaging techniques.  A noise ceiling
of~\SI{100}{(\meter\second^{-1})\squared\hertz^{-1}} was selected to
reject data above this level, and this has reduced the overall fill
from 82\,\% to 78\,\%.  The plot of data quality and noise levels
(Figure~\ref{fig:all_quality}) shows excellent stability through the
entire period.  Histograms of noise level and FOM
(Figures~\ref{fig:all_noisehist} and~\ref{fig:all_fomhist}) show the
daily distribution of data quality.

The new data pipeline provides an improvement across the entire
historic archive, not just new data, and as such provides an exciting
opportunity for new science.

%%%%%%%%%%%%%%%%%%%%%%%%%%%%%%%%%%%%%%%%%%%%%%%%%%%%%%%%%%%%%%%%

%% file: opendata.tex
% OPENDATA.TEX
%
%   Steven Hale
%   2015 May 27
%   Birmingham, UK
%
% BiSON Data Portal
%
% $Id: opendata.tex 49 2015-06-22 12:13:10Z hale $
%

%%%%%%%%%%%%%%%%%%%%%%%%%%%%%%%%%%%%%%%%%%%%%%%%%%%%%%%%%%%%%%%%

\section{BiSON Open Data Portal}
     \label{S-BiSON Open Data Portal}

%%%%%%%%%%%%%%%%%%%%%%%%%%%%%%%%%%%%%%%%%%%%%%%%%%%%%%%%%%%%%%%%

All data produced by the {BiSON} are freely available from the
\emph{{BiSON} Open Data Portal}~--~
\url{http://bison.ph.bham.ac.uk/opendata}
~--~and are also in the process of being deposited in the University
of Birmingham Long Term Storage Archive (LTSA).  The LTSA ensures the
data will be available via a persistent {URL} for a minimum of ten
years.  Data will be available from the archive using the
``FindIt@Bham'' service~--~
\url{http://findit.bham.ac.uk}
~--~and we also hope to provide all datasets with a Digital Object
Identifier (DOI) as soon as this facility becomes available via the
archive.

Data products are in the form of calibrated velocity residuals,
concatenated into a single time series from all {BiSON} sites.
Individual days of data, and also bespoke products produced from
requested time periods and sites, are available by contacting the
authors.  We hope to also provide all raw data products via the LTSA
as the archive is populated.  Oscillation mode frequencies and
amplitudes are available from~\cite{2009MNRAS.396L.100B}
and~\cite{2014MNRAS.439.2025D}.

All data created specifically during the research for this article are
openly available from the University of Birmingham ePapers data
archive \citep{epapers1977} and are also listed on the {BiSON} Open
Data Portal.

%%%%%%%%%%%%%%%%%%%%%%%%%%%%%%%%%%%%%%%%%%%%%%%%%%%%%%%%%%%%%%%%

%% file: future.tex
% FUTURE.TEX
%
%   Steven Hale
%   2015 May 27
%   Birmingham, UK
%
% BiSON Future
%
% $Id: future.tex 59 2015-10-17 06:44:36Z hale $
%

%%%%%%%%%%%%%%%%%%%%%%%%%%%%%%%%%%%%%%%%%%%%%%%%%%%%%%%%%%%%%%%%

\section{To The Future}
     \label{S-To The Future}

It would be scientifically advantageous to increase the network duty
cycle from 78\,\% to 100\,\% or better (\textit{i.e.} to have
every~\SI{40}{\second} interval covered by one or more sites), and
also to make better use of weighted averaging of multi-site data.  We
saw in Section~\ref{S-Site Performance} a seasonal variation in noise
level caused by the changing Doppler offset throughout the year.  A
similar offset effect is seen on a daily period due to Earth's
rotation, and this causes instruments at different longitudes to
sample line formation at different heights in the solar atmosphere
meaning that the noise between sites is not completely coherent.
Simultaneous observing at as many sites as possible allows the
incoherent components of the noise to be beaten down, and potentially
gives access to solar $g$-modes, which are expected to have very low
frequencies and low amplitudes~\citep{2010A&ARv..18..197A}.

Technology has moved on significantly since the {BiSON} nodes were
designed in the late 1980s.  Whilst a considerable programme of
upgrades has taken place over the years, the overall design is still
limited by the original specification.  Deploying more nodes in the
classic style would be prohibitively expensive.

By taking advantage of modern fiber optics, and electronic
miniaturisation such as micro-controllers and single-board computers,
it is possible to design a solar spectrometer with a much smaller
physical footprint and considerably lower deployment cost, thus making
it feasible to observe from many more sites.  Work is underway on a
second-generation network that will operate as a complement to the
existing nodes, and this aims to guarantee better than 100\,\% duty
cycle and much lower noise levels.
     
%%%%%%%%%%%%%%%%%%%%%%%%%%%%%%%%%%%%%%%%%%%%%%%%%%%%%%%%%%%%%%%%